\documentclass[a4paper, amsfonts, amssymb, amsmath, reprint, showkeys,
nofootinbib,onecolumn]{revtex4}
\usepackage[utf8]{inputenc}
\usepackage{color}
\usepackage[dvipsnames]{xcolor}

\usepackage{appendix}
\usepackage{soul}
\usepackage{ulem}
\usepackage{graphicx}
\usepackage{tikz}
\usepackage[pdftex, pdftitle={Article}, pdfauthor={Author}]{hyperref} % For hyperlinks in the PDF

\begin{document}

\title{ {Meson and glueball spectroscopy within the graviton soft wall 
model}}

\author{Matteo Rinaldi}
    \email[Correspondence email address: ]{matteo.rinaldi@pg.infn.it}% Your name
    \affiliation{Dipartimento di Fisica e Geologia. Università degli studi di
 Perugia. INFN section of Perugia. Via A. Pascoli, Perugia, 06123, Italy.}

\author{Vicente Vento}
\affiliation{Departamento de F\'{\i}sica Te\'orica-IFIC, Universidad de Valencia- CSIC,
46100 Burjassot (Valencia), Spain.}

\date{\today} % Leave empty to omit a date

\begin{abstract}

The graviton soft wall model (GSW) provides a unified description of the scalar
 glueball and meson spectra with a unique energy scale. 
This success has led us to extend the analysis to the description of the spectra
 of other hadrons. 
{We use this model to calculate masses of the}
 odd and even ground states of glueballs for various spins, and {
 show} that the 
GSW model is able to 
reproduce the Regge trajectory of these systems.
In addition,  the spectra of the $\rho$, $a_1$ and the $\eta$ mesons will
 be addressed. Results are in excellent agreement with current experimental 
data. Furthermore such an achievement is obtained without any additional 
parameters. Indeed, the only two parameters appearing in these
 spectra are those that were previously fixed  by the light scalar meson and 
glueball spectra.  
 Finally, in order to describe the   $\pi$ meson 
spectrum, a suitable modification of the dilaton profile function has been 
included in the analysis
 to properly take into  account  the 
Goldstone realization of chiral symmetry. The present investigation confirms that 
the GSW model provides an excellent  description of the spectra 
of mesons and glueballs with only a small number of parameters  unveiling a relevant predicting power.

\end{abstract}

\keywords{meson, glueball, gravity}

\maketitle

\section{Introduction}
In the last few years, hadronic models, inspired by  the holographic conjecture 
\cite{Maldacena:1997re,Witten:1998zw}, have been vastly used and developed in order 
 to investigate 
non-perturbative features of glueballs and mesons, thus trying to 
grasp fundamental features  of QCD 
\cite{Fritzsch:1973pi,Fritzsch:1975wn}. Recently we have 
used the so 
called 
AdS/QCD models to study the scalar glueball 
spectrum~\cite{Vento:2017ice,Rinaldi:2017wdn}. The holographic principle relies in a correspondence between a 
five dimensional 
classical theory with an AdS metric and a supersymmetric
 conformal quantum 
field theory with $N_C  \rightarrow \infty$. This theory, different from QCD, is taken as a starting point  to construct a 5 dimensional 
holographic dual of it. This is the so called bottom-up approach~\cite{Polchinski:2000uf,Brodsky:2003px,DaRold:2005mxj,Karch:2006pv}. 
In this scenario,  models are constructed by modifying  the five dimensional  
classical AdS theory with the aim of resembling  QCD as much as 
possible. The main differences characterizing these models are related to the 
strategy used to break conformal invariance.  Moreover, it must be noted that 
the relation which these models establish with QCD is at the level of the 
leading order in the number of colours expansion, and thus the mesonic and 
glueball spectrum and their decay properties are ideal observables to be studied 
by these models.

Being the mesons and glueball masses $\mathcal{O}(N_C^0)$, the AdS/QCD models reproduce the essential features of 
the meson and glueball spectrum~\cite{Erlich:2005qh,Colangelo:2008us,deTeramond:2005su,Rinaldi:2017wdn,Colangelo:2007pt,Capossoli:2015ywa}. 
For mesons and baryons, these approaches have been also successfully used to describe form factors and 
various types of parton distribution 
functions~\cite{deTeramond:2005su,Rinaldi:2017roc,Bacchetta:2017vzh,
deTeramond:2018ecg}. 
Besides these developments, which are in line with the 
present investigation, other models have been recently  introduced  by
using the
bottom-up holography. For example, an interesting development is the 
no-wall model 
\cite{Afonin:2013npa}, which has been successful in explaining the 
heavy meson spectra.

The present investigation has as its starting point  the holographic
Soft-Wall (SW) model scheme, were a dilaton field  
is introduced to softly break conformal invariance. This procedure allows to properly reproduce 
the Regge trajectories  of the meson spectra.
Within this scheme we have recently  introduced 
 the graviton soft-wall model 
(GSW)~\cite{Rinaldi:2017wdn,Rinaldi:2018yhf,Rinaldi:2020ssz} which
has been able to reproduce, not only the scalar meson spectrum, but also the 
lattice QCD scalar glueball 
masses~\cite{Morningstar:1999rf,Chen:2005mg,Lucini:2004my}, 
that was not described by the traditional SW models.  Moreover,  a formalism to 
study the glueball-meson mixing conditions {has been developed}  and  some 
predictions, regarding the observably of pure glueball 
states, {has been provided} ~\cite{Rinaldi:2018yhf,Rinaldi:2020ssz}.
 The success of the model  in reproducing the scalar QCD spectra, has motivated 
us to 
 extend, in the present investigation, the 
GSW model to describe the $\rho$ vector meson, the $a_1$ axial vector meson, the 
pseudo-scalar meson spectra and  to calculate the Regge trajectories of high 
spin glueballs. To develop a unified approach where the   QCD dynamics 
of glueballs 
is encoded in the modified metric, a specific dilaton, providing  the 
correct confining mechanism for a given hadron, is constructed.  To this aim, a 
differential equation for  the dilaton field has been 
obtained, which leads to
an effective phenomenological potential that  produces a good description of  
several meson spectra.

 In the next section,    the modification of the SW model, to obtain 
the GSW one, is discussed. In particular,  in Sect.~\ref{sec2} 
{we summarise the essence of the GSW model 
~\cite{Rinaldi:2017wdn,Rinaldi:2018yhf,Rinaldi:2020ssz}.
In Sect.~\ref{secglue} the GSW model has been applied to estimate 
the scalar 
glueball and spin dependent glueball spectra. In the last case the Regge 
trajectories have been  obtained and  they  compare 
successfully with lattice data.
In Sects.~\ref{sectscalar1}-\ref{sectdilato} the scalar spectrum is 
investigated. To this aim, a procedure to establish a dilaton which 
provides the 
correct confining mechanism has been developed. 
}
 In 
Sect. \ref{sec3}  the $\rho$ spectrum is described and in Sect. \ref{sec4} that 
of the $a_1$ meson will be shown together with a comparison with 
data.  In Sect. \ref{sec6} the pseudo-scalar 
meson spectra will be 
analysed and the GSW results presented. In Sect. \ref{conc} we discuss 
and summarise the results of our analysis.  {Finally, we have included 3 Appendices where the determination of the  dilaton equation is described in detail.}

\section{The GSW model}
\label{sec2}
Let us review in this section the essence of the GSW model. The development of this approach has been motivated by the impossibility of the conventional SW models to  describe
the glueball and meson spectra with the same energy 
scale~\cite{Rinaldi:2017wdn,Rinaldi:2018yhf,Rinaldi:2020ssz}. The essential 
feature, which distinguishes the GSW model from the traditional SW, is a 
deformation of the $AdS$ metric in 5 dimensions.

\begin{equation}
ds^2=\frac{R^2}{z^2} e^{\alpha \phi_0(z)} (dz^2 + \eta_{\mu \nu} 
dx^\mu dx^\nu) = e^{2 A(z)}  (dz^2 + \eta_{\mu \nu} 
dx^\mu dx^\nu) = e^{\alpha {\phi_0(z)}} g_{M N} dx^M 
dx^N  =\bar{g}_{MN}dx^M dx^N.
\label{metric5}
\end{equation}
where $A(z) = \log{{R}/{z} }+{\alpha \phi_0(z)}/{2}$.

The quantities evaluated in the GSW model will be displayed with overline. The 
function $\phi_0(z)$ will be specified later. This kind of 
modification has been adopted in many studies of the properties of mesons 
and glueballs within AdS/QCD 
~\cite{Colangelo:2007pt,Capossoli:2015ywa,Vega:2016gip,Akutagawa:2020yeo,
Gutsche:2019blp, Klebanov:2004ya,MartinContreras:2019kah,FolcoCapossoli:2019imm,
Bernardini:2016qit,Li:2013oda}. 
The relation between the standard $AdS_5$ metric and 
$\bar{g}_{MN}$ is

\begin{eqnarray}
\bar g^{MN} &= &e^{-\alpha \phi_0(z)} g^{MN}, \\
\sqrt{-\bar{g} } & = & e^{\frac{5}{2}\alpha \phi_0(z)} \sqrt{-g}.
\end{eqnarray}

Once the gravitational background has been defined by the model, the same 
strategy used in the SW case is considered to obtain the equations of 
motion for the different fields dual to given hadronic states. The action, in terms of the standard AdS metric of the  SW model,
is given by

\begin{align}
\label{prefactor}
 \bar S ={\int d^4xdz~e^{-\phi_0(z) \beta  }\sqrt{-\bar g} 
\mathcal{L}(x_\mu,z)}= \int d^4xdz~e^{\phi_0(z) 
\left(\frac{5}{2} \alpha -\beta+1  
\right) }\sqrt{-g} e^{-\phi_0(z)} \mathcal{L}(x_\mu,z)~,
\end{align}
 where here the prefactor $exp\left[ {\phi_0(z) \left(\frac{5}{2} \alpha 
+\beta+1\right)} \right]$ is due to the modification of the metric.   {The 
parameters $\alpha$ and  $\beta$ parametrise the internal 
dynamics of the hadrons of QCD in AdS, its holographic dual. In the AdS dynamics}, $\alpha$ characterises the 
modification of the metric, while $\beta$  characterises the SW model dilaton, 
namely the breaking of conformal invariance. If one considers, as a starting 
point, the GSW model as a modification of  the SW model, one is forced to fix 
$\beta$ to have the same 
kinematics~\cite{Rinaldi:2017wdn,Rinaldi:2018yhf,Rinaldi:2020ssz} which leads, 
in the case of scalar fields, to $\beta= \beta_s= 1+\frac{3}{2} \alpha$ 
and in the case of the vector { fields} to $\beta=\beta_\rho=
1+\frac{1}{2}\alpha$. The 
function $\mathcal{L}(x_\mu,z)$ is the Lagrangian density representing the 
hadronic system. 
{In Refs. 
\cite{Erlich:2005qh,Colangelo:2008us,deTeramond:2005su,Rinaldi:2017wdn,
Colangelo:2007pt,Capossoli:2015ywa}, the chosen dilaton profile 
function has been $\phi_0(z)= k^2 z^2$ .  {We start with the same dilaton, however in order to include the 
chiral symmetry behaviour of the pion,  this functional form of the dilaton has 
to be modified. Details 
will be discussed in Sect. VI. Moreover, as {it} will become clear in the 
next section, in order to 
properly describe confinement and thus the spectra, a further free 
parameter  addition to the dilaton has been introduced. }

{We proceed in the next section to describe the spectra 
of glueballs, and their relative Regge trajectories,  and 
we compare the results of our calculations with  lattice data}.

\section{GLUEBALLS IN THE GSW MODEL }
\label{secglue}

{ This section is dedicated to the successful application of the GSW model 
to the study of the glueball spectra and its comparison with lattice 
data.}

\subsection{Scalar glueballs as gravitons}
A peculiarity of the GSW model, which is the reason for the name, is that  the 
scalar glueball arises from the scalar component of the graviton and is not 
introduced as an independent field. Thus in our scheme the metric characterises 
the scalar graviton. Therefore the Einstein equation for the metric Eq. 
(\ref{metric5})  is the glueball mode equation.  In the 5th-variable $z$ once 
the $x$ dependence has been factorised  as $\Phi(z) e^{i x_\mu q^\mu}$, where 
$q^2 = -M^2$ and $M$ represents the mass of the glueball modes becomes,
 
 \begin{equation}
\label{7}
 \frac{d^2 \Phi(z)}{dz^2} - \left(\alpha k^2 z + \frac{3}{z}\right)
 \frac{d \Phi 
(z)}{d z} +  \left(\frac{8}{z^2} - 6 \alpha k^2 - 4 \alpha^2 k^4 z^2 + 
M^2\right) 
\Phi (z) - \frac{8}{z^2} e^{\alpha k^2 z^2} \Phi (z) = 0~.
 \end{equation}
By performing the change of function

\begin{equation}
\Phi (z) = e^{\alpha k^2 z^2/4} 
\left(\frac{z}{\alpha k}\right)^{\frac{3}{2}} \phi (z)
\end{equation}
we get  a Schr\"odinger type equation

\begin{equation}
- \frac{d^2 \phi(z)}{d z^2} + \left(  \frac{8}{z^2} e^{\alpha k^2 z^2} 
-\frac{15}{4} \alpha^2 k^4 z^2 +
7 \alpha k^2 - \frac{17}{4 z^2} 
\right) \phi(z) = M^2 \phi(z).
\end{equation}
In this equation it is apparent that $M^2$ represents the mode mass squared 
which will arise from the eigenvalues of a 
{\it Hamiltonian} operator scheme. 
It is convenient to move to the adimensional variable $ 
t=\sqrt{\alpha k^2/2}\; z$ and to define the mode by 
$\Lambda^2 = (2 /\alpha k^2)\; M^2$. The the equation becomes

\begin{equation}
-\frac{d^2 \phi(t)}{d t^2} + \left(\frac{8}{t^2} e^{2 t^2} - 15 t^2 + 14 - 
\frac{17}{4 t^2}  \right) \phi(t) = \Lambda^2 \phi(t).
\label{Gexact}
\end{equation}
This is a typical Schr\"odinger equation with no free parameters except 
for an energy scale in the mass determined by $\alpha k^2$. The 
potential term is uniquely determined by the metric and only the scale 
factor is unknown and will be determined from  lattice QCD.
This equation has no exact solutions but numerical  ones have been
found \cite{Rinaldi:2017wdn}. The above expression can be approximated by expanding}
the exponential up to second term to get a Kummer type equation. However, such a 
procedure does not lead to good results {and} the spectrum turns out to be 
too flat, see details on Refs. 
\cite{Rinaldi:2020ssz}. 
As one can see in the left panel of Fig. 
\ref{sgsm}, for the value $\alpha k^2\sim (0.37$ GeV)$^2$ the scalar linear 
glueball spectrum is well reproduced, see also Tab. \ref{Gmasses}.
Let us mention the recent study 
{SDTK~\cite{Sarantsev:2021ein,Klempt:2021nuf}} where the mass of the 
ground state of the scalar glueball has been extracted from a phenomenological 
analysis of the BESIII data of the $J/\Psi$ decays. The result obtained is 
very close to that predicted by the GSW model \cite{Rinaldi:2018yhf}.

\begin{table} [htb]
{\color{black}
\begin{center}
\begin{tabular} {|c |c| c| c |c |c |c|}
\hline
$J^{PC}$& $0^{++}$&$2^{++}$&$0^{++}$&$2^{++}$&$0^{++}$&$0^{++}$\\
\hline
MP & $1730 \pm 94$ & $2400 \pm122$ & $2670 \pm 222 $&  & &  \\
\hline
YC & $1719 \pm 94$ & $2390 \pm124$ &  &  &  &  \\
\hline
LTW & $1475 \pm 72$ & $2150 \pm 104$ & $2755 \pm 124$& $2880 \pm 
164 $& $3370
\pm 180$& $3990 \pm 277$  \\
\hline
SDTK  & $1865 \pm 25^{+10}_{-30} $  & & 
& & & \\
\hline
GSW  & $1920$ & $2371$ & $2830$& $2830$& 
$3289$& $3740$  \\
\hline
\end{tabular}  
\caption{Scalar glueball masses [MeV] from lattice calculations by MP
~\cite{Morningstar:1999rf}, YC~\cite{Chen:2005mg} and LTW 
~\cite{Lucini:2004my} and the recent analysis 
{SDTK~\cite{Sarantsev:2021ein,Klempt:2021nuf}}  together with the result 
of our calculation for $\sqrt{\alpha} k =370$ MeV, obtained by the GSW 
model~\cite{Rinaldi:2018yhf}.}
\label{Gmasses}
\end{center}
}
\end{table}

\subsection{High Spin Glueballs}
\label{sec5}
In order to describe even and odd high spin glueballs we follow the 
approach described in Refs. 
\cite{Capossoli:2015ywa,BoschiFilho:2005yh,FolcoCapossoli:2019imm}.
In this case the action, written in terms of pure $AdS_5$ five, is the same as 
that of the scalar case \cite{Rinaldi:2020ssz}:

\begin{align}
 \bar S = \int d^5x~\sqrt{-g}e^{-k^2z^2} \Big[ g^{MN}\partial_M G(x) \partial_N 
G(x)+ e^{\alpha k^2 z^2} M_5^2 R^2 G(x) \Big]~,
\end{align}

and therefore the equation of motion is obtained from: 

\begin{equation}
\label{glumot}
\partial_M(\sqrt{-g} e^{-\phi_0(z)} g^{MN} \partial_N G(x)) =  
\sqrt{-g}  e^{-\phi_0(z)(1-\alpha)}  M^2_{5}R^2  G(x).
\end{equation}

In order to describe a spin $J$ glueballs, 
one can add 
$J$ 
covariant derivatives in the gravity dual 
operator~
\cite{deTeramond:2005su,Capossoli:2015ywa,BoschiFilho:2005yh,
FolcoCapossoli:2019imm}. Therefore, for an even 
spin glueball, the operator has the form,

\begin{align}
 \mathcal{O}_{4+j}=F D_{\{\mu 1....}D_{\mu J \}}F,
\end{align}
which is a $p=0$ form whose conformal dimension  is $\Delta = 4+ J$. 
For the odd spin case, one considers the symmetrized operator,

\begin{align}
 \mathcal{O}_{6+j}=Sym Tr \Big(\tilde F_{\mu \nu}F D_{\{\mu 1....}D_{\mu 
J 
\}}F  \Big)~,
\end{align}
which is also a $p=0$ form whose conformal dimension $\Delta =  6 + J$.
By using, the relation between the conformal dimension and the mass in 
five dimensions Eq.(\ref{M5}), since the glueballs are  p-forms of index 
$p=0$, one gets that for even spin glueballs,

\begin{align}
\label{even}
 M_5^2R^2 &= J(J+4) \; \mbox{for even}~ J~;
\end{align}
and for the odd spin glueballs

\begin{align}
\label{odd}
M_5^2 R^2 &= (J+2)(J+6) \; \mbox{for odd}~J.
\end{align}

In this framework,  the EoM for the glueballs 
can be rearranged in a 
Schr\"odinger type equation:
 
\begin{align}
\label{glugen}
 -\psi''(z)+ \left[ \dfrac{B'(z)^2}{4} -\dfrac{B''(z)}{2}+ \dfrac{M_5^2 
R^2}{z^2}e^{\alpha k^2 z^2}  \right] 
\psi(z)=M^2 \psi(z)~.
\end{align}
{where,} 
$A(z)=\log(R/z)+{\alpha \phi_0}/{2}$ and $B(z)=-\phi(z)-3 
A(z)$, being again $\phi(z)= \beta_s k^2 z^2$.  
The above
 equation leads to

\begin{align}
 -\psi''(z)+ \left[ \left(\beta_s + \frac{3\alpha}{2} \right)^2k^4z^2 - 
2\left(\beta_s + \frac{3\alpha}{2}\right) k^2 + \dfrac{15}{4z^2} + 
\dfrac{M_5^2 
R^2}{z^2}
e^{\alpha k^2 z^2}  \right] \psi(z)=M^2 \psi(z)~.
\label{glueballeq}
\end{align}

In this case, since $M_5^2R^2 \geq 0$, see Eqs. 
(\ref{even}, \ref{odd}), the exponential term is positive and therefore 
the potential is binding. The exact
equation 
can be numerically solved for bound states. Results  of the  
calculations for the odd 
and even glueballs are shown in Table \ref{glub1} and \ref{glub2}, 
respectively, and will be discussed later.  {Let us recall that in our 
formalism for the scalars
$\beta_s=1+\frac{3}{2}\alpha$.}

\subsection{Odd glueballs}
Despite the lack of the data related to glueballs with $J \geq 1$ spin, 
several 
QCD lattice and model calculation are at our disposal 
~\cite{Morningstar:1999rf,Chen:2005mg,Meyer:2004gx,Gregory:2012hu,
LlanesEstrada:2005jf, Mathieu:2008pb,Szanyi:2019kkn, 
Szczepaniak:2003mr,Mathieu:2008bf,Landshoff:2001pp,Meyer:2004jc}.  
In order to evaluate this spectrum within the GSW model, Eq.  
(\ref{glueballeq}) should be solved  
to find the lowest mode corresponding to  $n=0$ and for $M_5^2 R^2 = 
(J+2)(J+6)$.
In  Table \ref{glub1} we compare the results of our calculations 
for the ground states with a series of lattice results and model 
calculations and we see that 
we obtain a quite  good agreement with them.
Also in this case let us remark that this is 
a parameter free calculation. Indeed, $\alpha$ and $k$, the only 
parameters of the model,  have been  fixed by the spectra of the 
scalar glueballs and light scalar mesons {\cite{Rinaldi:2020ssz}.
 The latter 
remark 
will be discussed   in the next section.  
It is important to stress that the present calculation is not a fit to the data 
but a direct evaluation of the spectrum without any free parameters.} From our 
results, shown in 
 Tab. \ref{glub1}, one can derive  the Regge trajectories:
{\color{black}
\begin{align}
J \sim 0.18 \pm 0.01M^2 - 0.75 \pm 0.28\;
\end{align}
where $M$  here is in GeV, result to be compared with that of Ref. 
\cite{LlanesEstrada:2005jf},

\begin{align}
 J \sim 0.18 M^2+0.25~.
\end{align}

\begin{table}[h]
\begin{tabular}{| c | c | c | c | c | c | c | c | c | c |}
\hline
$J^{PC}$ 
         & M\&P            & Ky  & My          & Ll &   Mta & Sz& This 
work  &
{Ref. \cite{Bernardini:2016qit}}& {Ref.
\cite{FolcoCapossoli:2019imm}} \\ \hline
$1^{--}$ & $3850 \pm 140$ & $3830 \pm 
130$ & $3240 \pm 480$ & $3950$ & $3990$ & 3001 & $3308 \pm 15$ &2400 &2630  \\ 
\hline
$3^{--}$ & $4130 \pm 290$ & $4200 \pm 
245$ & $4330 \pm 460$ & $4150$ & $4160$ &
$4416$ & $4451 \pm 12$   & 3030 &3700\\ \hline
$5^{--}$ &               &          
    &             & $5050$ & $5260$ &
$5498$  & $5752 \pm 10$   &5010 &4740\\ \hline
$7^{--}$ &            &           &          & $5900$ &     & & $6972 
\pm \;8$     &7000&5780 \\ \hline
\end{tabular}
\caption{Comparison of the masses of the ground states  for the odd spin 
glueballs (in MeV)  from the following sources,
  M\&P~\cite{Morningstar:1999rf}, Ky~\cite{Chen:2005mg},  
My~\cite{Meyer:2004gx}, Ll~\cite{LlanesEstrada:2005jf}, 
Mta~\cite{Mathieu:2008pb}, Sz~\cite{Szanyi:2019kkn}, 
  with the results of the GSW model {We also show results obtained by 
the models of Refs. \cite{FolcoCapossoli:2019imm,Bernardini:2016qit}.}} 
 
\label{glub1}
\end{table}

\subsection{Even glueballs}

We calculate here the spectrum for even spin glueballs  by 
means 
of  Eqs. (\ref{glueballeq}) and (\ref{even}) 
In this case, $M_5^2 R^2 = J(J+4)$. Let us remind that for this sector 
lattice 
data of both the ground and excited states for the $0^{++}$ and $2^{++}$ 
are available together with that 
of the ground states of $4^{++}$ and $6^{++}$ states. The interpretation 
of the spectrum of the $0^{++}$ was the motivation
behind the formulation of the GSW model 
~\cite{Rinaldi:2017wdn,Rinaldi:2018yhf,Rinaldi:2020ssz} and has been 
thoroughly studied, thus here we  {discuss} the behaviour of 
the ground state 
of the $2^{++},4^{++},6^{++}$ glueballs. As one can see in Tab. 
\ref{glub2}, the ground states are well reproduced. From the results 
shown in  Tab. \ref{glub2}, one can derive   the Regge trajectories,

\begin{table}[h]
\begin{tabular}{| c | c | c | c | c | c | c | c | c |}
\hline
$J^{PC}$  
         &M\&P  
   &Ky
            &My& Gy &    Sk & Mtb   & This work  & {Ref. 
\cite{FolcoCapossoli:2019imm}} \\ \hline
$2^{++}$ &$2400 \pm 145$ &$ 2390 \pm 150 $  & $2150 \pm
130$ &$2620\pm 50$ &$2420$ & $2590$ & $2695 \pm 21$&2080 
\\ \hline
$4^{++}$ &              &   
            &$3640 \pm 150$ & & $3990$ & $3770$ & $3920 \pm 14$&3170 \\ \hline
$6^{++}$ &           & 
            & $4360 \pm 460$ &  & & $4600$& $5141 \pm 12$&4220 \\ \hline
\end{tabular}
\caption{Comparison of the masses of the ground states  for the even 
spin glueballs (in MeV)  from the following sources,
  M\&P~\cite{Morningstar:1999rf}, Ky~\cite{Chen:2005mg},  
My~\cite{Meyer:2004gx}, Gy{\cite{Gregory:2012hu}, 
Sk~\cite{Szczepaniak:2003mr}, 
Mtb~\cite{Mathieu:2008bf},  
  with the results of the GSW model and that of the approach of Ref.
 \cite{FolcoCapossoli:2019imm}.}  }
\label{glub2}
\end{table}
{\color{black}

\begin{align}
 J \sim (0.21\pm0.01) M^2 + 0.58\pm 0.34, 
\end{align}
where $M$ is here in GeV. The slope is  in reasonable agreement with
 Refs. \cite{Landshoff:2001pp,Meyer:2004jc}, i.e., $0.25$.

In Table \ref{Gmasses} we show the lattice data the graviton solution (GSW) , the J = 0 solution GSW0 and the J = 2
solution for the same values of the parameters as before (GSW2). The graviton solution describes the data well, while the J = 0  and the J=2 solutions do not. One cannot try to justify the discrepancy in terms of the chosen energy scale since the J=0 solution would require a higher energy scale while the J=2 solution a lower. Somehow the graviton with its degeneracy between scalar and tensor seems to contain the appropriate physics.

\begin{table} [htb]
{\color{black}
\begin{center}
\begin{tabular} {|c| c| c| c| c | c | c |}
\hline
$J^{PC}$& $0^{++}$&$2^{++}$&$0^{++}$&$2^{++}$&$0^{++}$&$0^{++}$\\
\hline
MP & $1730 \pm 94$ & $2400 \pm122$ & $2670 \pm 222 $&  & &  \\
\hline
YC & $1719 \pm 94$ & $2390 \pm124$ &  &  &  &  \\
\hline
LTW & $1475 \pm 72$ & $2150 \pm 104$ & $2755 \pm 124$& $2880 \pm 
164 $& $3370
\pm 180$& $3990 \pm 277$  \\
\hline
SDTK & $1865 \pm 25^{+10}_{-30}$ & & & 
& &
\\
\hline
GSW& $ 1920$  & $ 2371$ & $2830 $ & $2830$&  $3289$  & $3740$ \\
\hline
GSW0 & $ 1411\pm 52$  &  & $1728 \pm 67 $ & &$1995\pm79 $ & $2231\pm 86$ 
\\
\hline
GSW2 & & $2695\pm 21 $& & $3179\pm 20$& & \\
\hline
\end{tabular}  
\caption{Glueball masses [MeV] from lattice calculations by MP
~\cite{Morningstar:1999rf}, YC~\cite{Chen:2005mg}, LTW 
~\cite{Lucini:2004my} and the recent analysis 
{SDTK~\cite{Sarantsev:2021ein,Klempt:2021nuf}} compared with the graviton 
(GSW) and field 
(GSW0 and GSW2) correspondences .}
\label{Gmasses}
\end{center}
}
\end{table}

\section{THE SCALAR MESON SPECTRUM }
\label{sectscalar1}
In this section we present the results of the calculations of the light and heavy scalar {meson}
spectra within the GSW model. 
In the light sector we get the following Equation of Motion (EoM)  
\cite{Rinaldi:2020ssz}), 

\begin{equation}
\partial_M(\sqrt{-g} e^{-\phi_0(z)} g^{MN} \partial_N S(x,z)) =  
\sqrt{-g}  e^{-\phi_0(z)(1-\alpha)}  M^2_{5}R^2  S(x,z).
\end{equation}
Once we separate the $x$ dependence by factorising $S(x,z) = \Sigma(z) e^{-i 
q_\mu x^\mu}$ with $q^2 = -M^2$, where  $M$ is the mass of the meson modes, 
we get

\begin{equation}
- \frac{d^2 \Sigma(z)}{dz^2} +\left(\frac{3}{z} + 2 k^2z\right) \frac{d 
\Sigma(z)}{dz} - \frac{M_5^2R^2}{z^2} e^{\alpha \phi_0(z)} \Sigma(z) = M^2 
\Sigma(z)~.
\end{equation}
By 
{recalling that} $\phi_0(z)= k^2 z^2 $ and 
performing the change of function 

\begin{equation}
\Sigma(z) =\left( \frac{z}{k}\right)^\frac{3}{2} e^{ k^2 z^2/2} 
\sigma(z),
\end{equation}
 a Schr\"odinger type equation can be obtained,

\begin{equation}
-\frac{d^2 \sigma(z)}{dz^2}+V_s(z) \sigma(z)=   -\frac{d^2 \sigma(z)}{dz^2} + 
\left( k^4z^2 + 2 k^2 + \frac{15}{4z^2} 
- \frac{3}{z^2} e^{\alpha k^2 z^2} \right) \sigma(z) = M^2 \sigma(z).
\label{smeq}
\end{equation}

For the scalar meson, the AdS mass is

\begin{equation}
M_5^2R^2 = (\Delta - p)(\Delta +p -4)
\label{M5}
\end{equation}
{where $\Delta$ is the conformal dimension and $p$ the p-form index. For the scalar field  $M_5^2 R^2=-3$ since the  $\Delta=3$ and $p=0$ \cite{Contreras:2018hbi}.}

{The potential  of the above equation, obtained with 
the same procedure used for the glueballs, is in this case not binding, 
 as shown by the full 
line of Fig. \ref{conf1}. Indeed, since for scalar mesons the conformal mass is negative, the 
exponential term in Eq. (\ref{smeq}) prevents the system to bind. However, 
in Ref. \cite{Rinaldi:2020ssz}, it has been shown that if the above potential 
 is Taylor expanded for small values of $\alpha$, up to the first three terms, 
a binding potential related to the  glueball dynamics of QCD described by the 
metric 
Eq. (\ref{metric5}), can be obtained. In the next section a formal procedure to 
motivate such a truncation will be presented together with its physical 
interpretation.}

\begin{figure*}
{\color{black}
\includegraphics[scale=0.80]{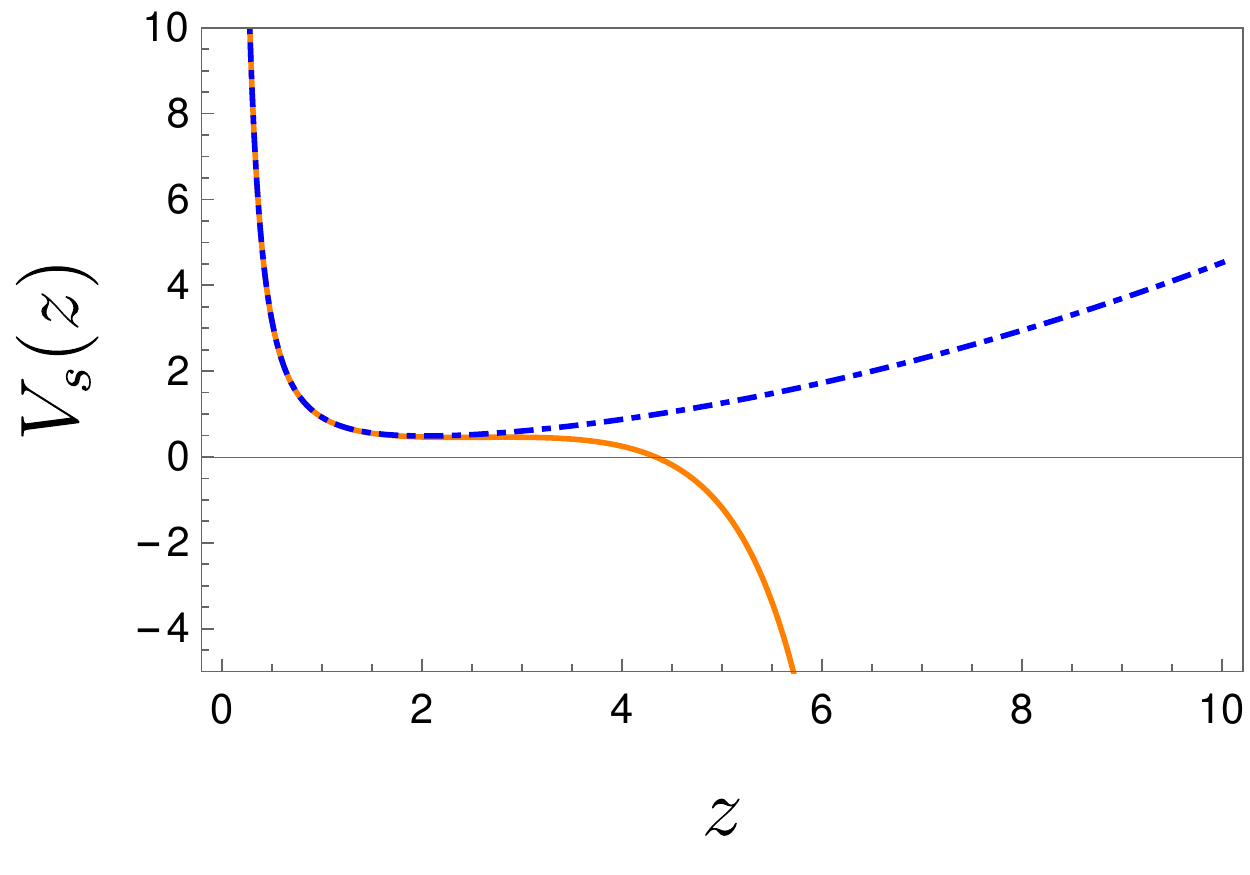} 
\caption{  The full potential in Eq. (\ref{smeq}) is shown by the solid line 
and 
the dot-dashed line shows the potential arising from the series expansion of the potential around small values of $\alpha$ keeping the first three terms. The figures shown have been  obtained for
 $\alpha = 0.55$.   }
\label{conf1}
}
\end{figure*}

\section{New dilaton and a phenomenological potential for mesons}
\label{sectdilato}
{In this section we present a new procedure to coherently describe the 
glueball  and the meson spectra within the  GSW model, i.e. the same metric. 
In fact, as previously discussed, at the variance with the spin dependent 
glueball, 
where the conformal mass is positive and thus the metric (\ref{metric5}) leads to a confining potential, 
in the meson sector, the relative potential is not binding due to negative the
 conformal mass. 
However,  it has been shown that if this 
potential is truncated after a 
Taylor 
expansion for small values of $\alpha$, the potential  confines and the 
spectrum is very well reproduced \cite{Rinaldi:2020ssz}. Therefore, in this 
section 
we show how the approximated potential  can be obtained.  Let us remark that 
the latter quantity is very appealing because, as discussed in Ref. 
\cite{Rinaldi:2020ssz}, it leads to a very good description of the light 
and heavy meson spectra with only one free parameter,  $\alpha$, since the scale 
parameter 
  $\alpha k^2$ was fixed by the 
spectrum of the scalar glueball \cite{Rinaldi:2017wdn}.  In this framework, we point out that  the 
procedure here introduced will not make use of any additional free parameter and the 
only  restriction consists in obtaining the convenient effective potential for 
the scalar meson 
EoM. Let us anticipate that, as it will be shown in the next sections, this 
type of potential allows to reproduce the spectra of various meson families without introducing any ad-hoc parameters.
To this aim, we consider a modification of the dilaton in the meson sector. 
In the following we consider the scalar case, however, as shown in Appendices 
\ref{appA}-\ref{appC}, the results can be
 generalised to the vector sector and to the pion, which will require a specific
 prescription to describe chiral-symmetry breaking. }
 
 Let us consider an extension of Eq. \ref{prefactor} for the scalar meson,   

\begin{align}
 \label{mod1}
\bar S = \int d^5x ~\sqrt{-g} e^{- \phi_ 0(z)
 -\phi_n(z)  } \Big[  g^{MN} 
\partial_M S(x) \partial_N S(x)+ e^{\alpha \phi_0(z)} M_5^2 R^2S^2(x) \Big]~,
\end{align}
{where we recall that  $\phi_0= \ k^2 z^2$. Furthermore,  we denote by 
$\phi_n$ an 
 addition to the dilaton $\phi_0$ with the purpose of generating the effective 
potential. The relative EoM is now,}

\begin{align}
 \label{eom4}
\partial_M \Big[\sqrt{-g} e^{-\phi_0(z) -\phi_n(z)   } g^{MN} \partial_N S(x)
\Big]-\sqrt{-g} M_5^2 R^2  e^{ \alpha \phi_0(z) -\phi_n(z) } S(x)=0~.
\end{align}
{Then potential in the corresponding Schr\"odinger  equation reads,  }

\begin{align}
V_s(z)=\frac{15}{4 z^2}+M_5^2 R^2  \frac{e^{\alpha k^2 z^2}}{z^2} +2 
k^2+k^4 
z^2 +\phi_n^{'}(z) \left( \frac{3}{2z} 
+ k^2 z  \right)  + \frac{\phi_n^{'}(z)^2}{4} 
-\frac{\phi_n^{''}(z)}{2}  ~.
\end{align}
{Now we compare the above potential with the one  obtained by  
considering the dilaton $\phi_0$ and the truncated exponential,}

\begin{align}
\label{potA}
V_s^A(z)=\frac{15}{4 z^2}+M_5^2 R^2  \frac{1 + \alpha k^2 z^2 + \frac{1}{2} 
\alpha^2 k^4 z^4}{z^2} +2 k^2+k^4 
z^2~.
\end{align}
{Out of this comparison we conclude that the addition to the old dilaton, $\phi_n$, is determined by solving the following second order differential equation,}

\begin{align}
-\frac{\phi_n^{''}(z)}{2} +\phi_n^{'}(z) \left( \frac{3}{2z} 
+ k^2 z  \right)  + \frac{\phi_n^{'}(z)^2}{4} 
+ \frac{M_5^2 R^2}{z^2} \left[ e^{\alpha k^2 z^2}-1-\alpha 
k^2 z^2- \frac{1}{2} \alpha^2 k^4 z^4  \right]=0~.
\label{difeqs}
\end{align}

{As one can see, the differential equation is highly non linear. However, 
a numerical solution can be found. In 
Fig. \ref{dilnew} we show the  evaluation of the dilaton together with 
a fit obtained by considering known profile functions. Further details are 
discussed in Appendix \ref{appA} where, a 
differential equation, valid for a scalar system, is shown without 
specifying 
the initial dilaton so that it can be applied to more general frameworks. 
Moreover, in Appendix \ref{appB} we have shown the equivalent expression for  
vector fields and, finally in Appendix \ref{appC} a general expression for the 
differential equation for the dilaton correction is found for both the scalar 
and the vector fields addressing the general behaviour for this addition.}

\begin{figure*}
{\color{black}
\includegraphics[scale=0.80]{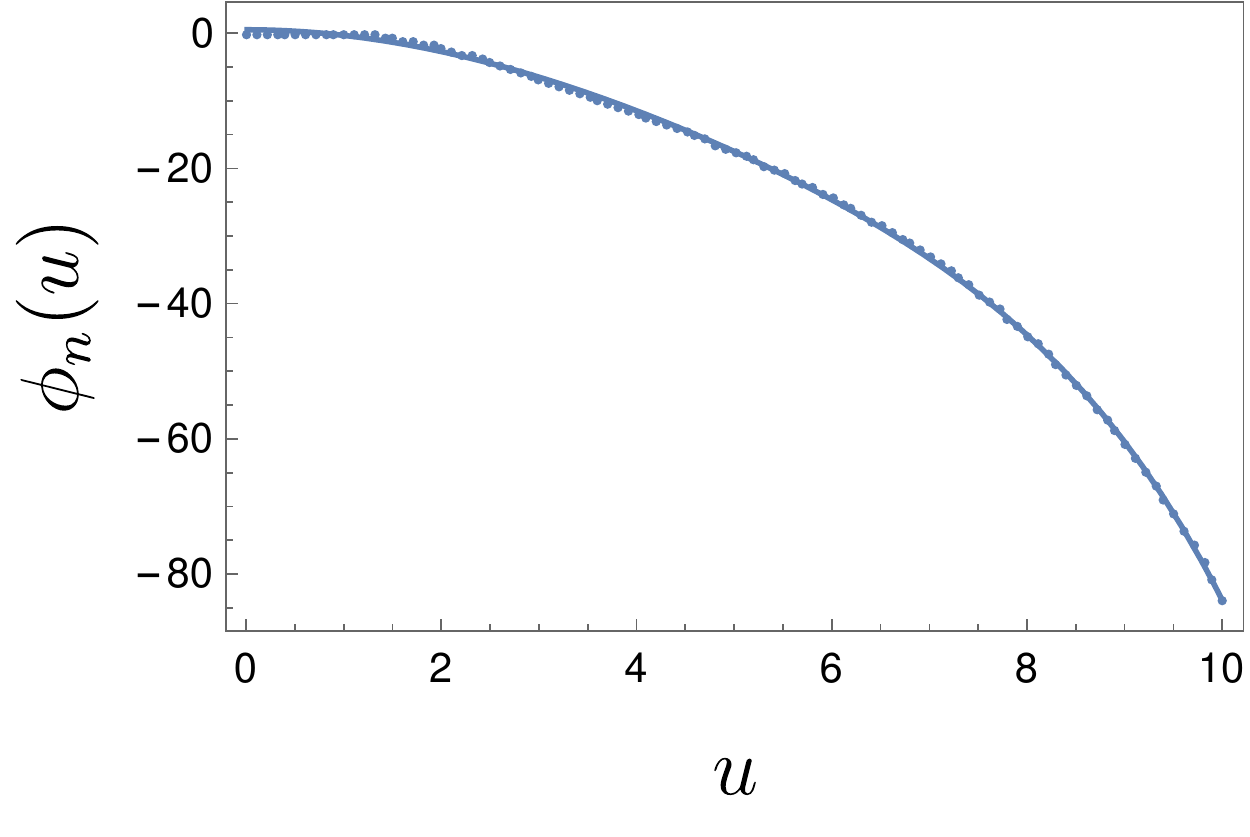} 
\caption{The dotted line shows the dilaton addition obtained by solving  numerically Eq. 
(\ref{difeqs}). The full line represents a fit to the dilaton addition obtained by the following 
function: $\phi_n(u) \sim a +b~ u^{3/2} +c~ u^2+d~ u^4+e ~u^6$ whose coefficients 
are shown in the Appendix \ref{appA}. Here $u= \alpha k^2 z^2$.  }
\label{dilnew}
}
\end{figure*}

{Once a solution has been shown to be exist, the equation of 
motion 
for the scalar meson is that obtained in Ref. \cite{Rinaldi:2020ssz} and with the 
potential shown in Eq.  
(\ref{potA}) which corresponds to the described truncation of the metric. 
Let us point out that for the moment being since
 we are mainly interested in the spectra,
the explicit expression of $\phi_n$
 is not needed once 
its existence is  verified.
 In closing this section we 
remark that the philosophy 
beyond the procedure is to reproduce a  phenomenological
potential which  leads
to an excellent description of the spectra despite its simplicity, as will be 
presented in the next sections. Furthermore, 
the 
procedure does not require any new free parameters making as clear as possible the 
physical interpretation. In fact, 
the dilaton is the 
mechanism used in the SW model to describe confinement. On the other 
hand,} the deformation of the 
metric  has been introduced to describe what in the dual
QCD sector is an additional interaction of gluons beyond confinement 
leading to the correct glueball spectra. {We} recall that we are 
dealing with  $1/N_c$ physics. Therefore, if this additional contribution 
destroys confinement in the meson sector, it cannot be correctly interpreted as 
a  {realistic} contribution to the SW model providing the correct 
binding energy for those systems. {Therefore, an appealing solution is to
 modify the  } dilaton, in the 
meson sector, to {dynamically compensate} the  metric {effects 
which prevent the binding}.  
{The consequent} truncation of  the exponential up to the third term 
{provides } confinement.  Thus, {such a procedure} can be  
physically interpreted as an attempt 
to estimate additional gluon effects beyond confinement  in the standard SW 
description of the mesons. {As it will be deeply shown later on,} the good 
results, in describing all the spectra, by using only two  
parameters, suggest that this procedure is appealing and 
realistic. {  In closing, let us stress again that despite the dependence 
of dilaton on the considered systems, as shown in Appendix \ref{appC}, the 
differential equation for the  addition $\phi_n$ has the same form 
 for 
scalar and vector fields. The differences arise due to their AdS mass $M_5^2R^2$ and  two calculated coefficients related to their kinematics (see  
Appendix \ref{appC} for details)}. Therefore we remark that  our procedure does 
not introduce any new freedom in the model.  We conclude that in the GSW 
model, confinement is determined by the interplay of the glueball dynamics of QCD described by the metric 
Eq. (\ref{metric5}), and { confinement described} by a well defined dilaton which leads to a 
phenomenological  binding potential. }

\subsection{The scalar meson with the new dilaton}
Motivated by the properties of the new dilaton, $\phi_0+\phi_n$, 
we recall here the main outcome of Ref. \cite{Rinaldi:2020ssz}, i.e. the light 
and heavy meson spectra 
 within the GSW model. 
As already discussed, the main effect of the correction $\phi_n$ is to 
produce a potential similar to that of Eq. (\ref{smeq}) but with the 
exponential 
truncated to the   third term. The final Schr\"odinger
equation Eq. (\ref{smeq}) is shown in terms of
 the adimensional variable $ u= 
\sqrt{k^2/2}\; z$,

\begin{equation}
-\frac{d^2 \sigma(u)}{du^2} + \left( 4 u^2 + 4 + \frac{15}{4u^2} - 
\frac{3}{u^2} e^{2 \alpha u^2} \right) \sigma(u) =  \Omega^2 \sigma(u),
\label{MAexact}
\end{equation}
where $\Omega^2= (2/k^2)M^2 $.

Expanding the exponential up to third order in Eq.(\ref{MAexact}) we get 

\begin{equation}
-\frac{d^2 \sigma(u)}{du^2} + \left(  (4 - 6 \alpha^{ \color{black} 2 }) u^2 + 
(4 - 6 \alpha) + 
\frac{3}{4u^2} \right) \sigma(u) =  \Omega^2 \sigma(u).
\label{Mapprox}
\end{equation}

This equation can be transformed into a Kummer type equation by the change of 
variables $v= (4- 6 \alpha^2)^{1/4}u$ 

\begin{equation}
-\frac{d^2 \sigma(v)}{d v^2} + \left(v^2 + \frac{4 - 6 \alpha}{\sqrt{4 - 6 
\alpha^2}} + \frac{3}{4 v^2}\right) \sigma(v) = \frac{\Omega^2}{\sqrt{4 - 6 
\alpha^2}} {\color{black}  \sigma(v)},
\label{Mkummer}
\end{equation}
which has an exact spectrum given by 

\begin{equation}
\Omega_n^2 = 4(n+1) \sqrt{4 - 6 \alpha^2} + 4 - 6\alpha, \; n = 
0,1,2,\ldots.
\label{Mmodes}
\end{equation}
and the mode functions are

\begin{equation}
\sigma(v) = \mathcal{N} e^{-v^2/2} v^{3/2} { _1F_1}(- n, 2, v^2)
\label{kummerf}
\end{equation}
where $\mathcal{N}$ is a normalisation factor and $_1F_1$ is a well 
known hypergeometric function and recall that $v = (4-6\alpha^2)^{(1/4)} \; u 
$
where $u = (\sqrt{k^2/2} \;z)$. Note that the approximate solution only has 
bound states for $|\alpha| < \sqrt{2/3}$. 
The meson modes  are  functions of $\alpha$.
As one can see in the left 
panel of Fig. \ref{sgsm} a good fit is found for $0.51 \leq \alpha \leq 
0.59$. 

\begin{figure}[h]
\includegraphics[scale=0.63]{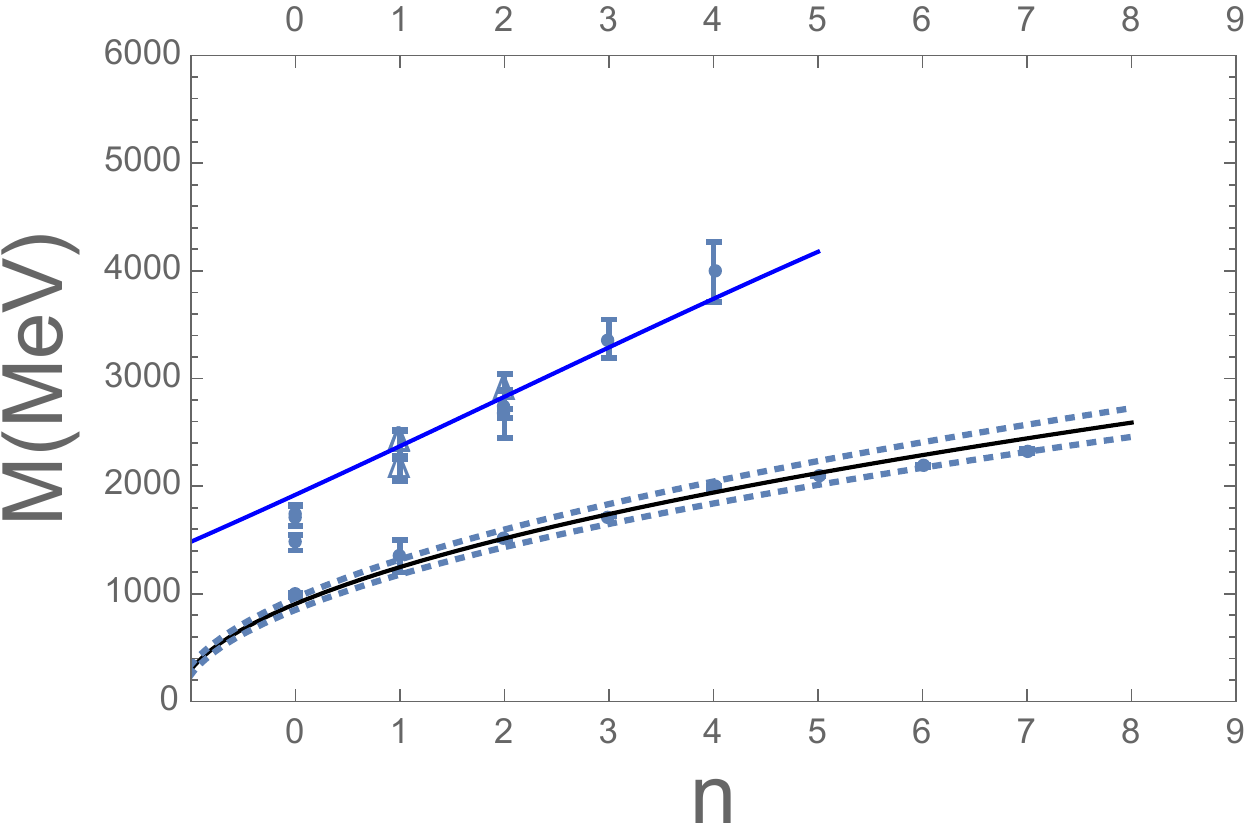} \hskip 0.2cm
\includegraphics[scale=0.65]{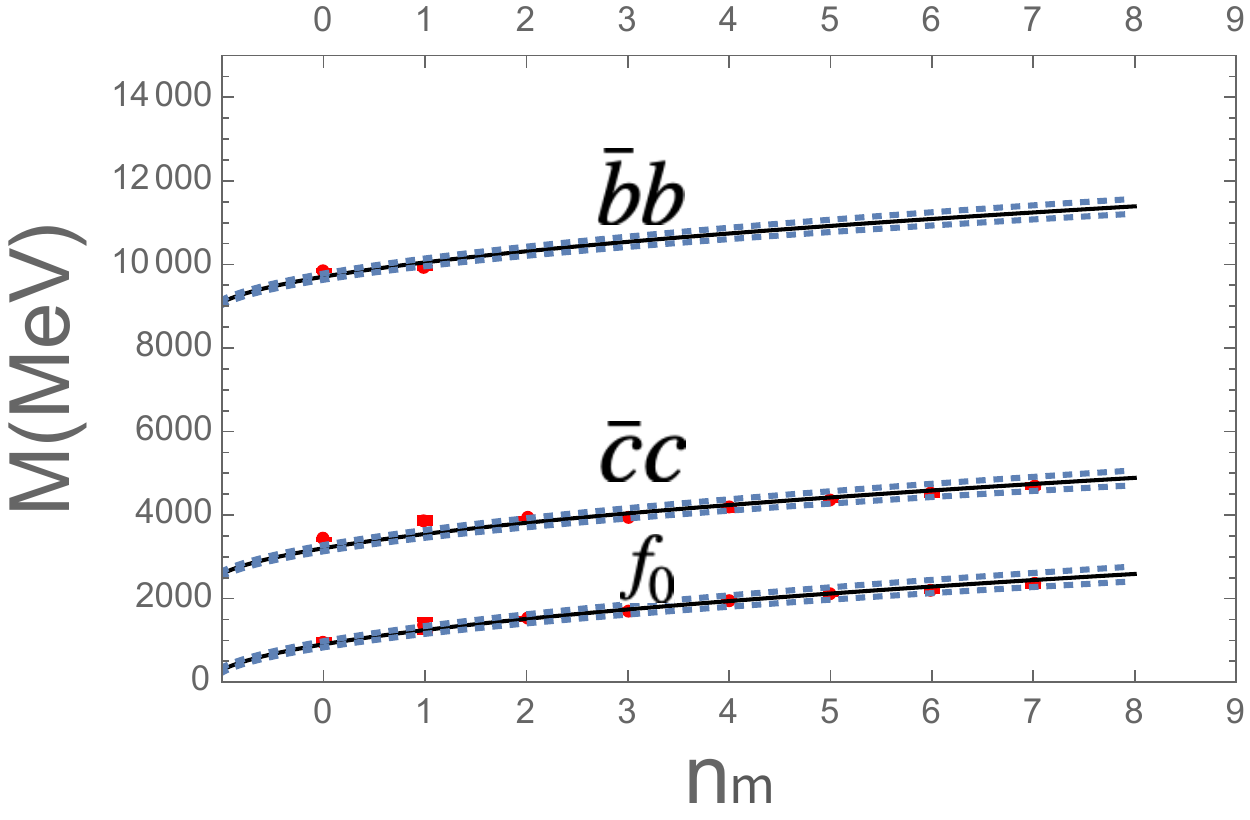}
\caption{\footnotesize  \textsl{
Left panel: GSW fit to the scalar lattice glueball 
spectrum~\cite{Morningstar:1999rf,Chen:2005mg,Lucini:2004my} and to the 
experimental scalar meson spectrum~\cite{Tanabashi:2018oca,Zyla:2020zbs}  .The 
larger slope of the glueball spectrum is noticeable. The  corresponds to $\alpha 
k^2=(0.37$GeV)$^2$ and the values of $\alpha$: $0.55$ (solid) and $0.55\pm0.04$ 
(dotted). Right panel: The scalar meson spectrum GSW fit to the data  shown for 
all quark sectors. The 
light dots represent the scalar meson spectrum experimental data 
~\cite{Tanabashi:2018oca,Zyla:2020zbs}. The  curves correspond to 
Eq. (\ref{heavymass}) with $C_c = 2400$ MeV for the $c \bar{c}$ mesons 
and $C_b = 8700$ MeV for the $b \bar{b}$ mesons and the values of  $\alpha$: 
$0.55$ (solid) and $0.55\pm0.04$ (dotted) for all 
mesons. }}
\label{sgsm}
\end{figure}

\subsection{Heavy mesons}

In addition in Ref. \cite{Rinaldi:2020ssz} it has been shown that the GSW can 
also reproduce the heavy meson spectra by following the procedures developed in 
Refs. \cite{Branz:2010ub,Kim:2007rt,Afonin:2013npa}, i.e. by including in the 
dynamics the mass of the heavy quarks.
Among the different possibilities, we have used  the following 
ansatz:

\begin{equation}
M_n = \sqrt{A(\alpha) n + B(\alpha)} + C,
\label{heavymass}
\end{equation}
where $C$ is the contribution of the quark masses, thus there will be a 
$C_c$ for the $c \bar{c}$ states and a different one $C_b$ for
the $ b \bar{b}$ states. The comparison between data 
~\cite{Tanabashi:2018oca,Zyla:2020zbs} and 
predictions are shown in the right panel of Fig. \ref{sgsm}.
In order to perform the calculation, the value of  $\alpha$ has been kept fixed 
from that obtained in the analysis of light mesons, i.e.
$\alpha=0.55$ (solid) and $\alpha=0.55\pm0.04$ (dotted). Moreover,  
 $C=0$ for the light quark sector,   
$C_c=2400$ MeV, for the $ c \bar{c}$ mesons, and for the $ b \bar{b} $ 
mesons $C_b=8700$ MeV. As one can see in the right panel of Fig. \ref{sgsm}, 
the model reproduces the data extremely well. Moreover, one should notice that 
the 
additional parameters $C_c$ and $C_b$ are extremely close to the value of $2 
m_c$ and $2 m_b$, respectively, as expected. Let us stress  that the heavy quark 
sector has not been considered to 
estimate the value of $\alpha$. From the  present calculations one can conclude 
that  
all  mesons satisfy approximately  the same mass trajectories apart from an 
overall scale associated 
with the quark masses and that all the elements in Refs. 
~\cite{Tanabashi:2018oca,Zyla:2020zbs} 
suspected of being scalar mesons seem to be scalar mesons, except for some 
possible mixing with the low lying scalar glueballs, which is not contemplated 
in this  scheme \cite{Rinaldi:2020ssz}. The model has proven to be tremendously 
predictive.
{Details on the comparison with data are displayed in Tab. 
\ref{heavymasstable}}. 
We summarise this section by stating that  
the GSW model describes well the scalar lattice glueball and the 
phenomenological scalar meson spectra of QCD with only two parameters, i.e. 
$\alpha$ and energy scale $\sqrt{\alpha}\, 
k$~\cite{Rinaldi:2017wdn,Rinaldi:2020ssz}.

\begin{table} [htb]
\begin{center}
\begin{tabular} {|c |c |c |c |c |c |c |c |c|}
\hline
light & $f_0(500)$ 
&$f_0(980)$&$f_0(1370)$&$f_0(1500)$&$f_0(1710)$&$f_0(2020)$&$f_0(2100)$&$f_0(
2200)$\\
\hline
$I^G(J^{PC})$& 
$0^+(0^{++})$&$0^+(0^{++})$&$0^+(0^{++})$&$0^+(0^{++})$&$0^+(0^{++})$&$0^+(0^{++
})$&$0^+(0^{++})$ &$0^+(0^{++})$\\
\hline
PDG & $475 \pm 75$ & $990 \pm 20$ & $1350 \pm 150 $&$1504 \pm 6 $  & $1723 \pm 6 
$&  $1992 \pm 16 $&  $2101 \pm 7 $&$ 2189 \pm 13$\\\hline
 GSW model \cite{Rinaldi:2020ssz} & $ $ & $907 \pm 73$ & $1248 \pm 93 
$&$1514 \pm 109 $  & $1740\pm 123 $&  $1941 \pm 136 $&  $2121 \pm 147 $&$ 2288 
\pm 157$\\\hline\hline
$c \bar{c} $ &$ 
\chi_{c0}(1P)$&$\chi_{c0}(3860)$&$X(3915)$&$X(3940)$&$X(4160)$&$X(4350)$& 
$\chi_{c0}(4500)$&$\chi_{c0}(4700)$
\\
\hline
$I^G(J^{PC})$& 
$0^+(0^{++})$&$0^+(0^{++})$&$0^+(0/2^{++})$&$?^?(?^{??})$&$?^?(?^{??})$&$0^+(?^{
??})$&$0^+(0^{++})$ &$0^+(0^{++})$\\
\hline
 PDG & $3414 \pm 0.30$ & $3862^{+66}_{-45}$ & $3918 \pm 1.9 $&   
$3942^{+13}_{-12}  $&  $4156^{+40}_{-35} $&  $4350^{+5.3}_{-5.1} $ & 
$4506^{+42}_{-41}  $ & $4704^{+24}_{-34} $\\\hline
 {\color{black} GSW model} \cite{Rinaldi:2020ssz} & $3307 \pm 73$ & $3648 \pm 
93$ & $3914 
\pm 109 $&$4141 \pm 123 $  & $4340 \pm 136 $&  $4521\pm 147 $&  $4688 \pm 157 
$&$ 4844 \pm 168$\\\hline\hline
 $b \bar{b}$& $ \chi_{b0}(1P)$&$ \chi_{b0}(1P)$& & & & & & \\
\hline
$I^G(J^{PC})$& $0^+(0^{++})$&$0^+(0^{++})$& & & & & &  \\
\hline
 PDG & $9859 \pm 0.73$ & $9912.21 \pm 0.57$ &  &  &  & & & \\
\hline
{\color{black} GSW model}  \cite{Rinaldi:2020ssz} & $9707 \pm 73$ & $10048 \pm 
92$ &  &  
&  & & & \\
\hline

\end{tabular}  
\caption{Scalar meson spectrum [MeV] from the PDG listings 
~\cite{Tanabashi:2018oca,Zyla:2020zbs} together with our results for 
$\alpha=0.55\pm 0.04$ and the energy scale $\sqrt{\alpha}\, k =370$ MeV, see 
Ref. \cite{Rinaldi:2020ssz} for details. Notice that in the PDG listings some of 
the particles are only suspected to be scalars and others need confirmation.}
\label{heavymasstable}
\end{center}
\end{table}

\section{The $\rho$ vector meson spectrum}
\label{sec3}
Let us apply the GSW model to the calculation of the spectrum of the vector meson family of the $\rho$.  We consider a vector field in the 
modified $AdS$ space. 
The respective action~\cite{FolcoCapossoli:2019imm}, 
modified with the GSW metric, reads,

\begin{align}
  \bar S = - \dfrac{1}{2} \int d^5x~ \sqrt{-\bar g}~ e^{\beta_\rho k^2 z^2} 
\left[ \dfrac{1}{2} \bar g^{MP} \bar g^{QN} F_{MN} F_{PQ} + M_5^2 R^2 \bar 
g^{PM} A_P A_M  \right]~,
\end{align}
where for the rho meson the AdS mass given by Eq.(\ref{M5}) leads to $M_5^2  R^2
=0$ since the conformal dimension is $\Delta=3$ and the p-form index $p=1$ 
\cite{Contreras:2018hbi}. 
{Since $M_5^2 R^2=0$, there is no need to add a correction to the initial 
dilaton $\phi(z)=\beta_\rho k^2 z^2$, in fact, the simplest solution to the 
relative differential equation, see Appendix \ref{appB}, is $\phi_n(z)=0$.}
Therefore, as previously discussed, if one moves to the standard  
AdS metric, the action can be rearranged as,

\begin{align}
 \bar S = - \dfrac{1}{2} \int d^5x~ \sqrt{-g}~ e^{ k^2 z^2 (\beta_\rho 
+\alpha/2 )} 
\left[ \dfrac{1}{2}  g^{MP}  g^{QN} F_{MN} F_{PQ}  \right]~.
\end{align}
{\color{black} Let us remark that in this case the GSW model is formally 
equivalent to 
the SW one because the deformed metric  does not affect the EoM since  $M_5^2 R^2 
=0$. Nevertheless, we anticipate that the energy scale $k$ will not be 
considered {as} a free parameter but instead we will use the value fixed 
in the scalar sector.}
As {discussed in the previous section,} we fix  $\beta_\rho$ 
 by  
imposing  that the kinematic term in the action is the same as that in the 
usual  SW $AdS$ action, thus $\beta_\rho = 1+\frac{1}{2}\alpha$. After this choice 
the action becomes,

\begin{align}
 \bar S = - \dfrac{1}{2} \int d^5x~ \sqrt{-g}~ e^{ -k^2 z^2 } 
\left[ \dfrac{1}{2}  g^{MP}  g^{QN} F_{MN} F_{PQ}  \right]~,
\end{align}
which is the same expression used in Ref. \cite{FolcoCapossoli:2019imm}.  Also in this 
case, an EoM
 in the Schr\"odinger form can be found:

\begin{align}
 -\psi''(z)+ \left[ \dfrac{B'(z)^2}{4} -\dfrac{B''(z)}{2} \right] 
\psi(z)=M^2 \psi(z)~.
\end{align}
{By setting $B(z)= k^2 z^2 + \log{z}$, the above expression becomes:}

\begin{align}
 -\psi''(z)+ \left(\frac{3}{4 z^2}+ k^4 z^2\right) \psi(z) =M^2 \psi(z)~,
\end{align}
This equation can be exactly solved  and the spectrum is given by

\begin{align}
 M_\rho = 2 \sqrt{1+n}~ k~.
\end{align}
Due to the fact that the five dimensional mass is zero, this formula coincides 
with that given in  Ref. \cite{Karch:2006pv} however now we have no freedom to 
fix the parameter $k$ which is given by 
 $k=370/\sqrt{\alpha}$ MeV with  $\alpha$ determined from the scalar glueball 
and meson spectra to be $0.51 \leq \alpha \leq 0.59$.  
{One should notice} that we can mathematically recover  the results of Ref. 
\cite{FolcoCapossoli:2019imm} by setting  $\beta_\rho=0$ and $\alpha=1$.
 
\begin{figure*}
\includegraphics[scale=0.80]{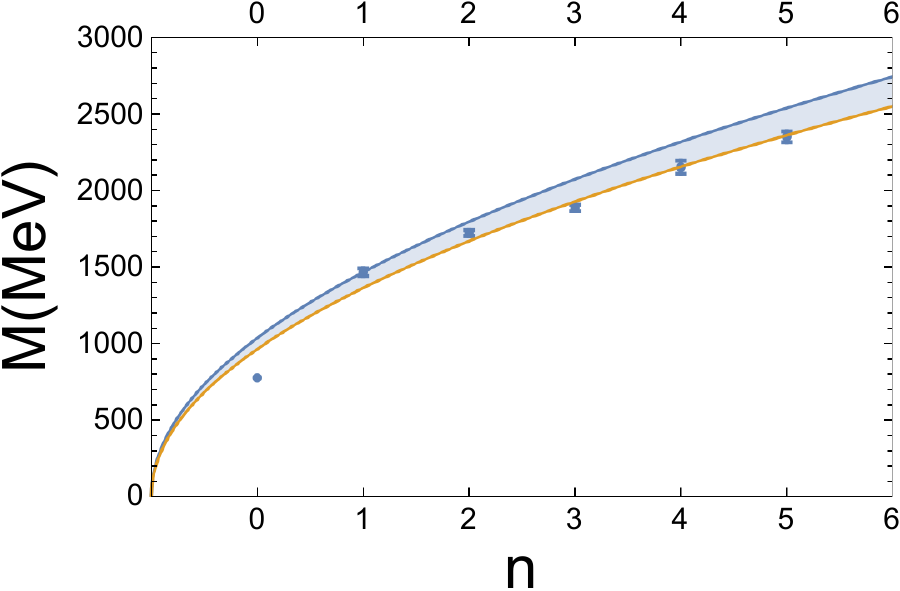} \hskip 1cm \includegraphics[scale=0.80]{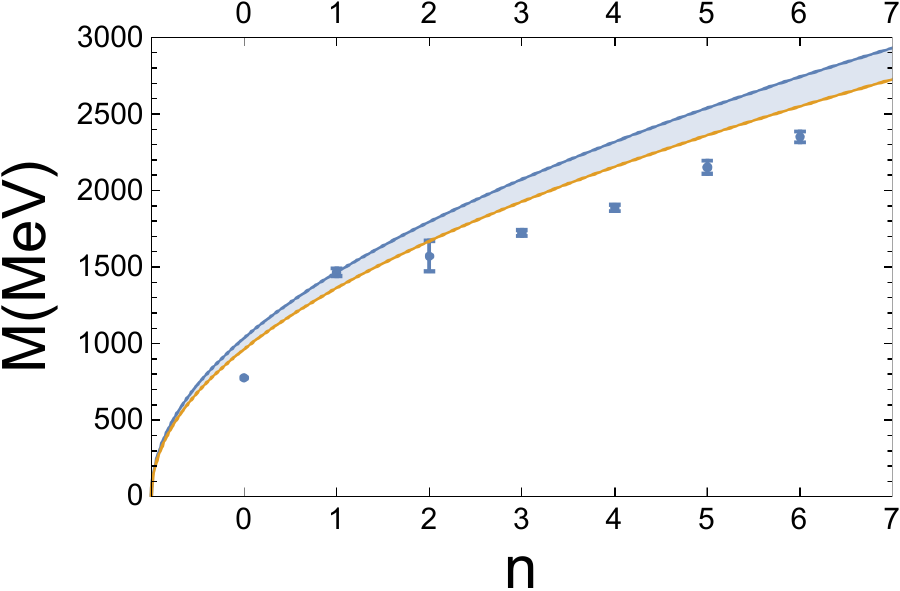} 
\caption{Left: The $\rho$ mass plot as a function of mode number according to the GSW model compared with the data where the experimental $\rho(1570)$ has been excluded following the discussion of the PDG particle listings ~\cite{Tanabashi:2018oca,Zyla:2020zbs}. Right: the same plot with the $\rho(1570)$ included. The result is not a fit since the parameters have been determined from the scalar mesons and the scalar glueballs.}
\label{rho}
\end{figure*}

\begin{table}[htb]
\begin{tabular}{| c | c | c | c | c | c | c | c | c |}
\hline
& $\rho(770)$ &$ \rho(1450)$ & $\rho(1570)$ & $\rho(1700)$& $\rho(1900)$  & $\rho(2150) $ &$ \rho(2350) $\\ \hline
PDG&$ 775.26\pm 0.25$&$1465\pm25$ & $1570\pm36\pm62$ & $1720\pm20$ & $1885 \pm 22$ & $2151 \pm 42$&$ 2330 \pm 35$ \\ \hline
This work &   $997\pm 38$   &$ 1411\pm 54$&    & $ 1728\pm 66 $ &$ 1995\pm 76$& $2231\pm 85$ &$2444 \pm 94$ \\ \hline
{Work of Ref. \cite{FolcoCapossoli:2019imm}} &   $868.3$   &$ 1228$&     $ 
1504 $ &$ 1736.7$& $1941.6$ & 2127 &\\ \hline
\end{tabular}
\caption{ We show the experimental result for the $\rho$ masses in MeV together 
with the results of our calculation for $\alpha=0.55\pm0.04$ and 
$k=370/\sqrt{\alpha} $ MeV. The experimental values were taken from the PDG 
particle listings ~\cite{Tanabashi:2018oca,Zyla:2020zbs}. {We also report 
the results obtained by the model of Ref. \cite{FolcoCapossoli:2019imm}.} 
}\label{rhot}
\end{table}

The phenomenological $\rho$ spectrum requires some comments before we show our 
results. {The} $\rho$ mesons are characterized by 
$J^{P\,C} = 1^{- \,-}$. However, looking deeper into the phenomenological 
analysis ~\cite{Tanabashi:2018oca,Zyla:2020zbs} the $\rho(1570)$ is supposed to 
be  OZI forbidden decay of the $\rho(1700)$ and therefore is not a pure $\rho$ 
state. 
As one can see in the left plot of  Fig. \ref{rho} and Table \ref{rhot} we get 
an overall good result for the spectrum. We must stress that our result is not a 
fit since we have taken the parameters from the scalar  {sector}. 
In the right  {panel of of  Fig. \ref{rho}} we 
include the $\rho(1570)$ as the $n=3$ mode to show that this incorporation 
distorts completely the agreement. Thus the GSW model 
{predicts}  that the $\rho=1700$ is the  $n=3$ mode and not the 
$\rho=1570$.  Some authors take out the so called D-rho 
mesons from the S-rho mesons in the mass fit~\cite{FolcoCapossoli:2019imm}. 
{Since the GSW model is well defined in the large $N$ limit, the present approach can not distinguish the above states.}
Our mode number $n$ acts as a good quantum number and incorporates S and D 
states. Our result shown in Fig. \ref{rho} and Table \ref{rhot} reproduces all 
the masses of the rho meson states with the precision  required from a large $N$ 
approximation. The main discrepancy,  {i.e.} the $\rho(770)$, 
 has to do with the observation that the low lying strongly bound 
states are not so well reproduced in large $N$ QCD.

 If one represents $M^2$ as a function of  $n$ one gets straight lines whose 
slope is $4 k^2 = 4 \times 0.37^2/\alpha$  which is in the range 
 {$1.002 \pm 0.072$}, included in  the universal 
range  $1.25 \pm 0.25$ \cite{Anisovich:2000kxa}. The difference comes again from 
the discrepancy in the mass of the $\rho(770)$.

\section{The $a_1$ axial meson spectrum}
\label{sec4}

\begin{table}[htb]
\begin{tabular}{| c | c | c | c | c | c | c | }
\hline
& $a_1(1260)$ &$ a_1(1420)$ & $a_1(1640)$ & $a_1(1930)$ & $a_1(2095)$& $a_1(2270)$ \\ \hline
PDG \& Av& $ 1230\pm 40$&$1411^{+15}_{-13}$ & $1655\pm16$ & $1930^{+19}_{-70} $  & $2096^{+17}_{-121}$ & $2270^{+55}_{-40}$ \\ \hline
This work &  833 $ \pm 53 $    &$1235 \pm  72$&       $ 1535   \pm 87$& $  1785 \pm 100$ &$ 2005\pm 111$& $2202\pm 122$\\ \hline
{Work of Ref. \cite{Contreras:2018hbi}} &  808.1 $  $    &$1114.7$&       $ 
1351.3$& $  1558.7$ &$ 1744.3$& $1913.4$\\ \hline
\end{tabular}
\caption{ We show the experimental result for the $a_1$ masses in MeV together 
with the results of our calculation for $\alpha=0.55\pm0.04$ and 
$k=370/\sqrt{\alpha} $ MeV. The three lower masses were taken from the PDG 
particle listings ~\cite{Tanabashi:2018oca,Zyla:2020zbs} and the three higher 
masses from Ref.~\cite{Anisovich:2001pn}. A comparison with the results 
obtained by the model of Ref. \cite{Contreras:2018hbi} is also shown.}
\label{a1t}
\end{table}

In the present approach,
the only difference between vector mesons and  axial-vector mesons due to 
chiral symmetry breaking is that the latter have $M_5^2R^2 \neq 
0$, see Ref. \cite{Huang:2007fv} for details. A mechanism for chiral symmetry 
breaking  {can} change the mass equation by introducing an anomalous 
conformal dimension $\Delta_p$ \cite{Contreras:2018hbi}

\begin{equation}
M_5^2R^2 = (\Delta + \Delta_p - p)(\Delta +\Delta_p +p -4).
\label{M5A}
\end{equation}
$\Delta_p=0$ for scalar mesons and vector mesons and turns out 
to be $\Delta_p=-1$ for pseudoscalar mesons and axial vector mesons.
 Thus $M_5^2R^2 = -4$ for pseudoscalars and $M_5^2R^2 = -1$ 
for axial vector mesons. Therefore the EoM for the $a_1$ becomes

\begin{align}
 -\psi''(z)+ \left(\frac{3}{4 z^2}+ k^4 z^2 - \frac{e^{\alpha k^2 z^2} }{z^2}  
\right) \psi(z) =M^2 \psi(z)~.
\end{align}
{As already discussed several times, such a potential is not binding. 
Therefore also in this case a modification of the dilaton, $\phi_n$, is 
included so that the effective potential is obtained by expanding the  term 
$e^{\alpha k^2 z^2}$ in the above expression up to the second order. The 
differential equation for $\phi_n$  is explicitly shown in the Appendix 
\ref{appB}~. The corresponding spectrum equation reads:}

\begin{align}
M^2 =  \big[4 (n+1) \sqrt{1-\alpha^2}- \alpha \big]  k^2,
\end{align}
which naturally for $\alpha=0$ coincides with the vector meson mass. Using our 
fixed value $k=370/\sqrt{\alpha}$ MeV and $\alpha=0.55\pm0.04$ we get the 
spectrum shown in Fig.\ref{a1} and Table \ref{a1t}, which is  close to the 
experimental results. Our calculation favors that the $ a_1(1930), 
a_1(2095)$ and $a_1(2270)$, not appearing in the PDG 
~\cite{Tanabashi:2018oca,Zyla:2020zbs} but shown in Ref. 
\cite{Anisovich:2001pn}, are axial resonances, thus favoring more 
experimental research in those energy regions.
{Moreover,  in the right panel of Fig. \ref{a1}, the 
same data are shown shifted by one unit in $n$. As one can see the agreement of 
the 
calculation with data increases. Therefore, in this very first application of 
the GSW model to the axial vector spectrum, one could propose that the model 
predicts the existence of a missing ground state with a mass lower then the 
quoted 1230 MeV.  }

\begin{figure*}[htb]
\includegraphics[scale=0.6]{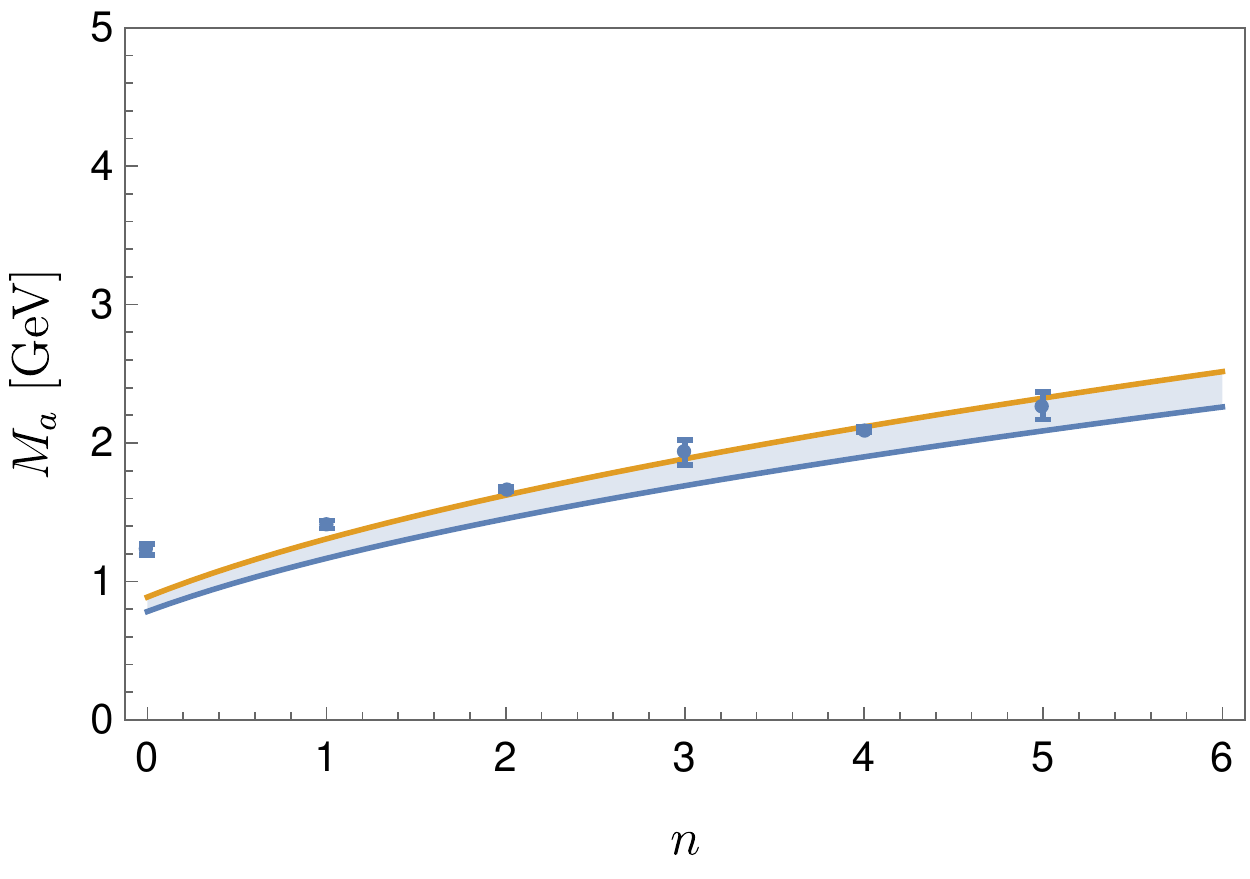}
\hskip 0.1cm
\includegraphics[scale=0.6]{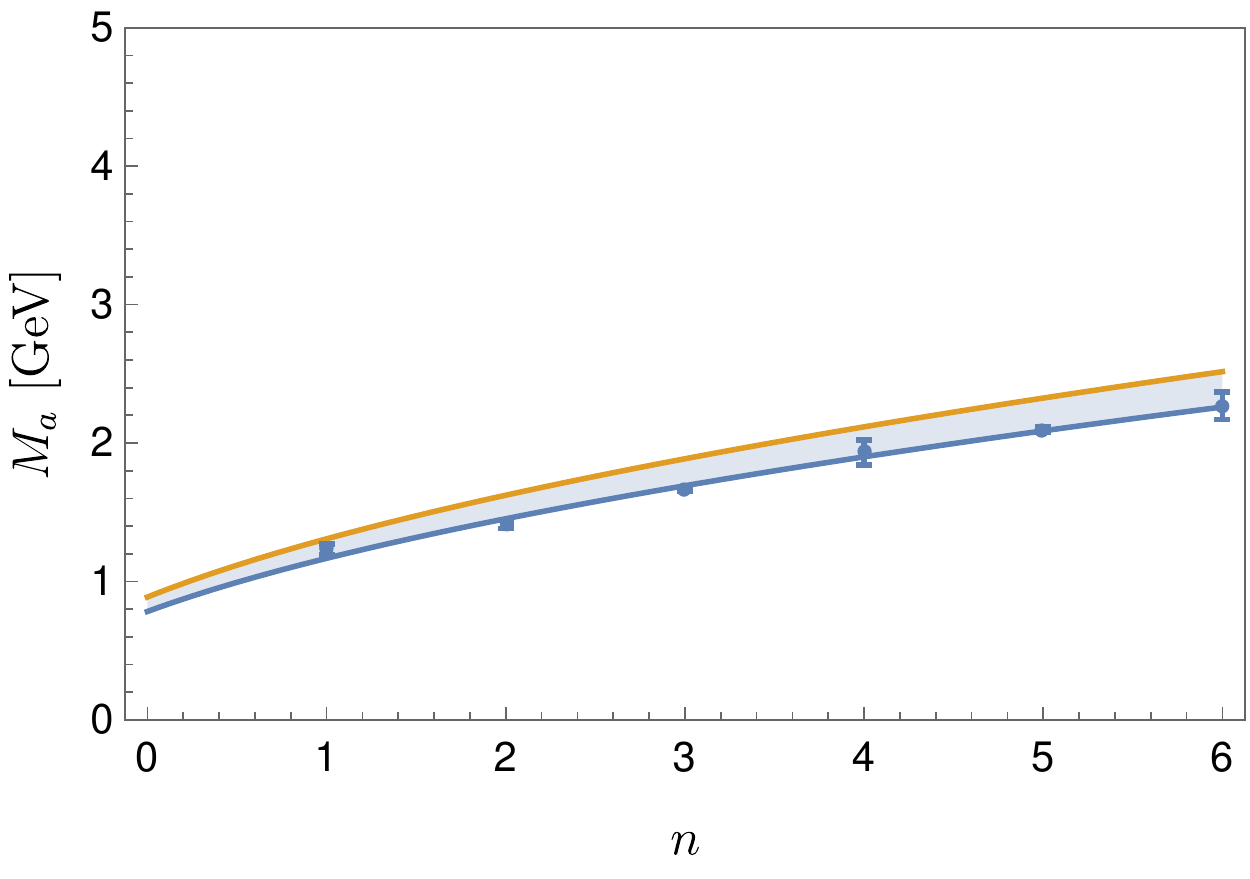}
\caption{ The $a_1$ mass plot as a function of mode number according to the GSW 
model compared with the data. The three lighter $a_1$ experimental masses are 
from Ref.~\cite{Tanabashi:2018oca,Zyla:2020zbs}, while the three heavier ones 
have been taken from Ref.\cite{Anisovich:2001pn}. The result is not a fit since 
the parameters have been determined from the scalar mesons and the scalar 
glueballs.
 Left panel:
calculation vs data. Right panel: same of the left one but data shifted in the 
$n$ axis by one unity.}
\label{a1}
\end{figure*}

\section{Pseudo-scalar mesons}
\label{sec6}

Lastly we will discuss the spectrum of the pseudo-scalar mesons. The EoM is governed by the conformal dimension 
related to dual field operator of the considered hadron. For a pseudo-scalar meson the study of the conformal dimensions leads to an AdS mass $M_5^2 R^2 = -4$ \cite{Contreras:2018hbi}. Thus we are going to consider the spectrum of particles characterized by $J^{PC} =0^{- +}$, which correspond in the spectroscopic notation,  previously used, to $J=0$ and $L$=0.  In this case the EoM becomes

\begin{equation}
-\frac{d^2 \psi(z)}{dz^2} + \left( k^4z^2 + 2 k^2 + \frac{15}{4z^2} 
- \frac{4}{z^2} e^{\alpha k^2 z^2} \right) \psi(z) = M^2 \psi(z).
\label{psmeq}
\end{equation}

{Since in this approach the pseudo-scalar and scalar mesons are described 
within the same formalism, the only different is $M_5^2 R^2$ and therefore 
one can add the correction of the dilaton which satisfies Eq. (\ref{difeqs}). 
Thus,
the truncated potential is recovered.}
From the phenomenological point of 
view there are two families of pseudoscalar particles: the $\pi$'s and the 
$\eta$'s. Let us first study the $\eta$'s.

\begin{figure*}[htb]
\includegraphics[scale=0.9]{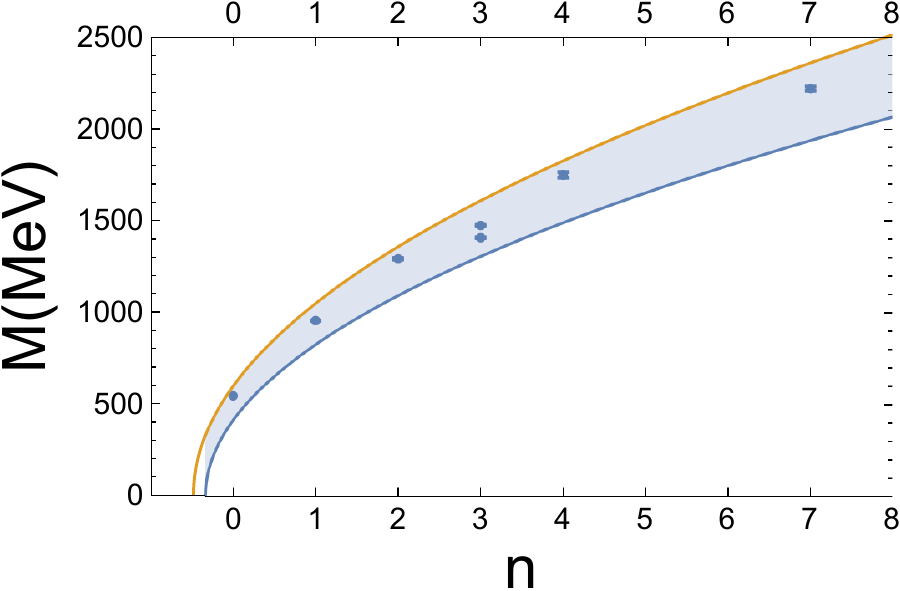}
\caption{The spectrum of the $\eta$ meson. The upper curve corresponds to  $\alpha=0.51$ and the lower curve to $\alpha=0.59$. We have given $\eta(1405)$ and $\eta(1475)$ the same mode number, $n=3$, (see discussion in the text), and we have skipped modes $n= 5, 6$ since the mass gap between the $\eta(1760)$ and the $\eta(2225)$ is double that of the lower mass $\eta$'s.}
\label{etaMass}
\end{figure*}

\subsection{The $\eta$ pseudoscalar meson}

\begin{table}[htb]
\begin{tabular}{| c | c | c | c | c | c | c | c | c |}
\hline
& $\eta$ &$\eta^{\prime}$ & $\eta(1295)$ & $\eta(1405)$ - $\eta(1475)$& $\eta(1760)$& $\eta(????) $  &$ \eta(????) $  &$\eta(2225)$ \\ \hline
PDG & $ 547.862\pm 0.017$&$957.78\pm0.06$ & $1295 \pm 4$ & $1408.8 \pm 2.0 $  & $1751 \pm15$ &&& $2221\pm 12$ \\
&&&&$1475 \pm 4$ &&& &\\ \hline
This work &   $ 513\pm 92 $    &$943 \pm  111$&       $ 1231   \pm 133$& $  1463 \pm 151$ &$1663 \pm 168$& $1842\pm 183$&$2005\pm198$&$2155 \pm 210$\\ \hline
\end{tabular}
\caption{ We show the experimental results for the $\eta$ masses, in MeV,  given by the PDG reviews ~\cite{Tanabashi:2018oca,Zyla:2020zbs} together with the results of our calculation. The gaps are introduced in order to respect the mass gaps of the GSW model calculation. }
\label{etatab}
\end{table}

 In Fig. \ref{etaMass} we show our calculation where the band  characterizes 
$\alpha= 0.55\pm 0.04$. In Table \ref{etatab} we show the PDG values of the 
$\eta$ masses ~\cite{Tanabashi:2018oca,Zyla:2020zbs} compared with the results 
of our calculation. It is discussed in PDG review that the $\eta(1405)$ and the 
$\eta(1475)$ might be the same particle, which is what seems to indicate our 
calculation. Moreover in the upper mass sector the experimental mass gap becomes 
larger, which according to the GSW model might indicate that some eta resonances 
are experimentally missing. In the Table  \ref{etatab} and in Fig. \ref{etaMass} 
we have left those mode numbers empty  between the $\eta(1760)$ and the 
$\eta(2225)$. If one trusts the results of the calculation, the GSW model 
predicts the existence of two resonances between the $\eta (1760)$ and the 
$\eta(2225)$ and that the $\eta(1405)$ and $\eta(1470)$ seem to be the same 
resonance. From the flavour content it is know that the $\eta$'s have hidden 
strangeness and therefore corrections associated to the quark mass should be 
added. Since the strange quark  is not too  heavy the corrections will be 
smaller than our theoretical errors. We must stress, once more, that this 
calculation is not a fit since our parameters were fixed by the scalar 
spectrum.

\subsection{The $\pi$ pseudoscalar spectrum}
 The main difference
between the $\eta$ and the $\pi$ is the isospin, however since our model does 
not take into account Coulomb corrections, the charge-less pions behave very 
much like the $\eta$ from the point of view of quantum numbers and therefore the 
spectrum  {should be} the same in our model but it is not  so in nature. In fact, there is one 
main difference, indeed the pion is  the Goldstone boson of SU(2) 
x SU(2) chiral symmetry and this fact is instrumental in giving the lightest 
pion its low mass.  {We should therefore implement the spontaneous broken 
realization of chiral symmetry in the  GSW 
holographic model to reproduce the low mass of the 
ground state pion.} 
The physics of confinement and chiral symmetry breaking is described by
 the dilaton ~\cite{Erlich:2005qh,Gherghetta:2009ac,Vega:2016gip}.
{Therefore, to the present aim, a modification of the dilaton profile function is here proposed}
 to 
implement the realization of chiral symmetry in a phenomenological way. The 
dilaton besides the conventional behaviour $\phi_0(z) \sim \beta_s k^2 z^2$ at 
large $z$, which determines Regge behaviour, requires a different behaviour  at 
low $z$ to implement chiral symmetry, namely $\varphi(z) \sim 
\beta^{\prime}k^2 z^2+\mathcal{O}(z^4)$ , with $\beta^{\prime} < \beta$ 
\cite{Gherghetta:2009ac}.
 {To the present aim,}  an efficient 
{choice for the dilaton profile function is to promote} 
$\beta$ to be a function of $z$

\begin{align}
\beta(z) = \beta_{\infty} \tanh{(\gamma z^4 + \delta)}~.
\end{align}

{This kind of ansatz has been considered several times in different 
analyses to implement the chiral symmetry breaking in holographic models, see 
Refs. \cite{Erlich:2005qh,Gherghetta:2009ac,Vega:2016gip}. This ansatz leads, as required by chiral symmetry, to,}

\begin{align}
\lim_{z\rightarrow \infty} \beta(z) = \beta_{\infty}~,
\end{align}
{and}
\begin{align}
\lim_{z\rightarrow 0} \beta(z) = \beta_{0} = \beta_{\infty} \tanh{(\delta)} + \mathcal{O}(z^4) < \beta_{\infty}.
\end{align}

 {The term}    
$ \tanh{(\delta)}$ is therefore related to the realization of chiral symmetry. With this 
phenomenological input, the dilaton function becomes

{
\begin{align}
\label{49}
\phi_0(z) = \beta_{\infty} \tanh{(\gamma z^4 +\delta)} k^2 z^2
\end{align}
}
with $\beta_{\infty}=\beta_s =1+\frac{3}{2}\alpha$ to satisfy the correct 
large $z$ behaviour once we take into account the effect of the GSW 
metric. Thus, 
the large $z$ behaviour, which dominates the spectrum of the higher modes, leads 
to the Regge behaviour. In the low $z$ region, which 
{is related to}  the transition region,  $\delta$ and $\gamma$, characterise 
{the}  spontaneous chiral symmetry breaking beyond $\Delta_p$, 
{i.e.,} the effect associated with the bulk 5D mass discussed previously. 
With this new dilaton, the equations of motion can be generated  {by using the  
same strategy previously discussed but }
{by} introducing the new dilaton in the functions $B(z)$. {As one 
might expect, the relative potential is more complicated w.r.t. the $\eta$ and 
other mesons. We again perform an expansion of the exponential 
to 
keep the largest binding potential and dismiss the terms which make it not 
confining.}
 {From a phenomenological point of view, one should expect that the value of 
$\gamma$ depends on the hadron under scrutiny. }
In particular, in the case of the low mass pion,  $\gamma$ must be 
relatively low so that the transition to the large $z$ limit occurs at higher 
values. Let us try to fit the low mass pion with the new dilaton. In 
Fig.\ref{pionnew} we show the wave function of a $135$ MeV pion for 
$\delta=1.5325$ and $\gamma=0.0055$ GeV$^{-4}$. The value of $\delta$ has 
been chosen to have a $n=0$ mode in the approximate well behaved solution. We 
have kept this value fixed in the non linear full equation and have varied only 
the $\gamma$ to get a solution as close as possible to the approximate one. 
The other 
parameters $\alpha$ and $k$ have been fixed as above.

\begin{figure*}[htb]
\includegraphics[scale=0.7]{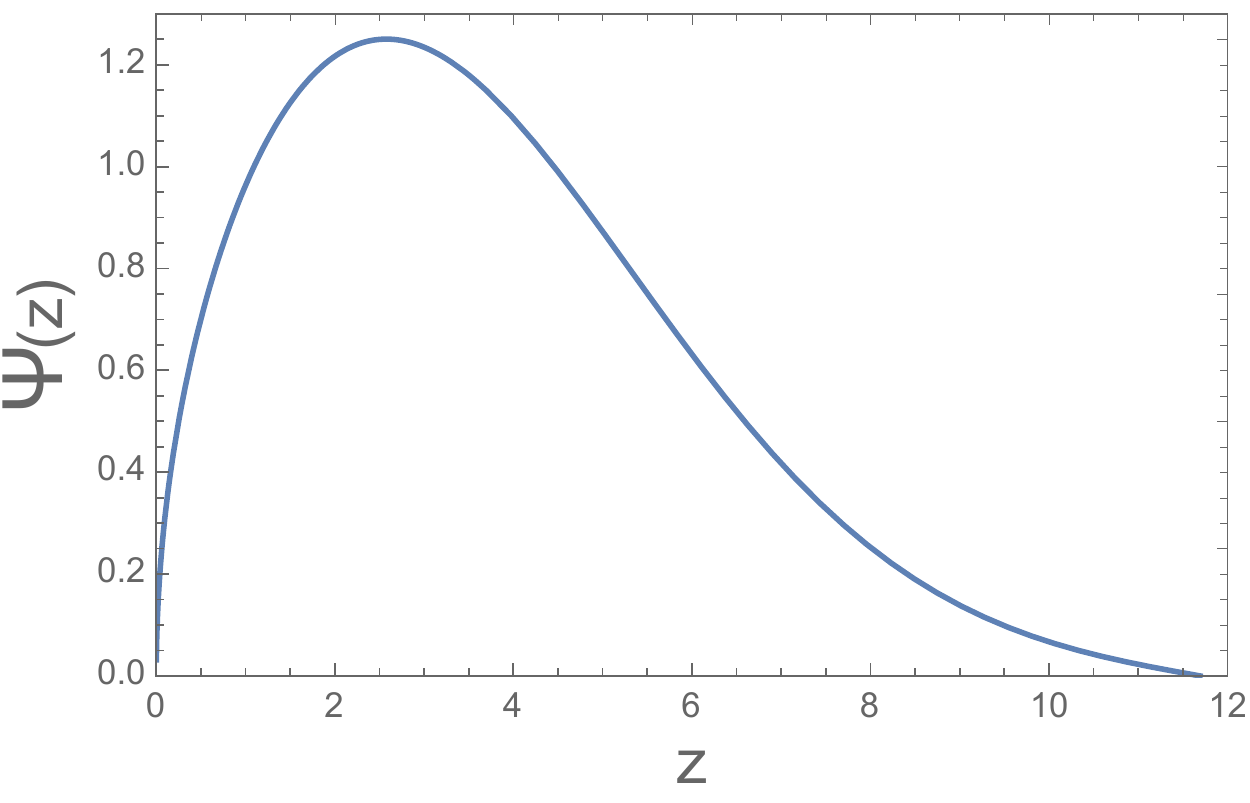}
\caption{We show the wave function for a $135$ MeV pion with the new dilaton.
 We have used  $\delta=1.5235$ and $\gamma =0.0055$GeV$^{-4}$, $\alpha=0.55$ and 
$k=0.37/\sqrt{\alpha}$ GeV .}
\label{pionnew}
\end{figure*}

{In order to calculate the full pion spectrum, one should notice that the 
excited states, namely $n \geq 1$, are not Goldstone bosons, therefore it is 
reasonable to assume that the relative EoM should be that described by Eq. 
(\ref{psmeq}). In other words, within this prescription, the underline dynamics 
generating resonances of the pion should be similar to that of the $\eta$ 
meson.}
{The full pion spectrum is displayed} in the Table \ref{piontab} 
compared with the PDG data~\cite{Tanabashi:2018oca,Zyla:2020zbs}. 
{As one might notice, the GSW model, incorporating the chiral symmetry 
breaking effect in the dilaton profile function, predicts a number of
pion states bigger than those experimentally observed. Such  a feature is shared 
with other models \cite{Contreras:2018hbi,Gherghetta:2009ac}. However, let us remark that the present experimental 
results  are somehow not conclusive. Indeed, the $\pi(1300)$ has a not well 
defined mass and a large width (over $200$ MeV), thus it might hide two 
resonances within its huge width. Moreover, the $\pi(1800) $ has experimental results 
ranging from $1770$ to $1870$ and also a large width and a complicated 
two pic 
structure.
In this scenario, the large $N$ approximation, encoded in approach
 here proposed, is predicting states that could be observed
 once the experimental region is cleared up.  }

 Finally one may wonder if the new dilaton will change the
 {$\eta$} spectrum, {previously} described. Let us show that this is 
not the case.  In Fig. \ref{etanew}, the wave functions of the first 
two $\eta$ modes, for the masses determined above $512$ and $943$ MeV with the 
new dilaton (solid) and the old dilaton (dashed), {are shown}. One sees that 
keeping fixed $\delta=1.5235$, which should be the same for all particles, and 
letting $\gamma$ grow up  to
 $\gamma \sim 1.0$, to displace the transition region to lower $z$ values, 
 one obtains exactly the same spectrum and almost exactly the 
same wave functions. For larger values of $\gamma$ the resemblance of the wave functions 
is even greater, but then we are mathematically approximating the old 
dilaton mode. For higher modes, the required value of $\gamma$, for almost 
equality, can be even lower since the $z^4$ term dominates the $\beta(z)$ 
function. For values as low as $ \gamma=0.2$, the wave functions are not so 
close but still very similar. 
{Furthermore, as a cross check of the procedure to provide a binding 
potential, in Fig. \ref{piondil} it has been numerically shown that the 
correction to the dilaton applied to the $\eta$ case is very close to that needed 
for the $\pi$ case. Such a feature reflects that also in the GSW model, the 
$\chi$sb can be described by a new dilaton $\phi_0$ (\ref{49}), while the 
truncation of the metric effect in the potential can still be obtained by means 
of  corrections that for the pseudo-scalar systems are very similar.  }

\begin{figure*}[htb]
\includegraphics[scale=0.65]{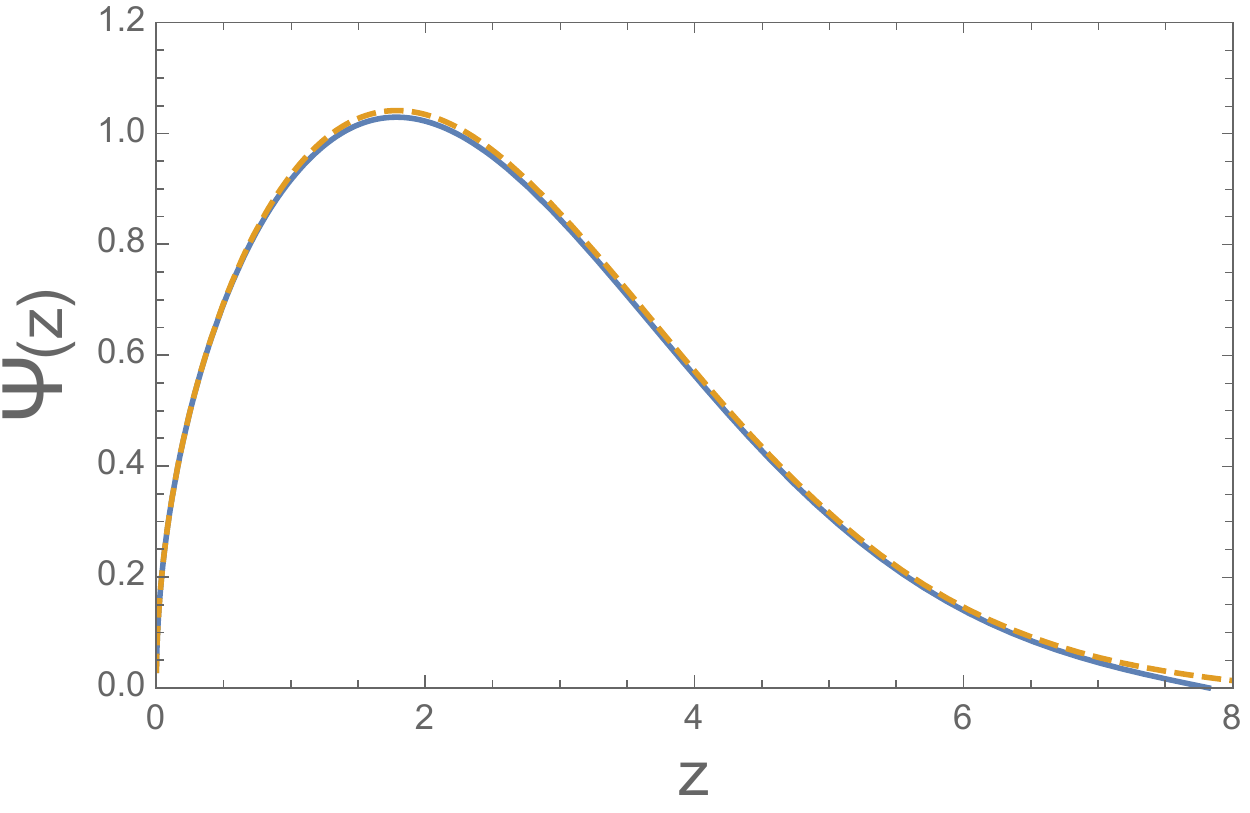} \hskip 0.7cm \includegraphics[scale=0.65]{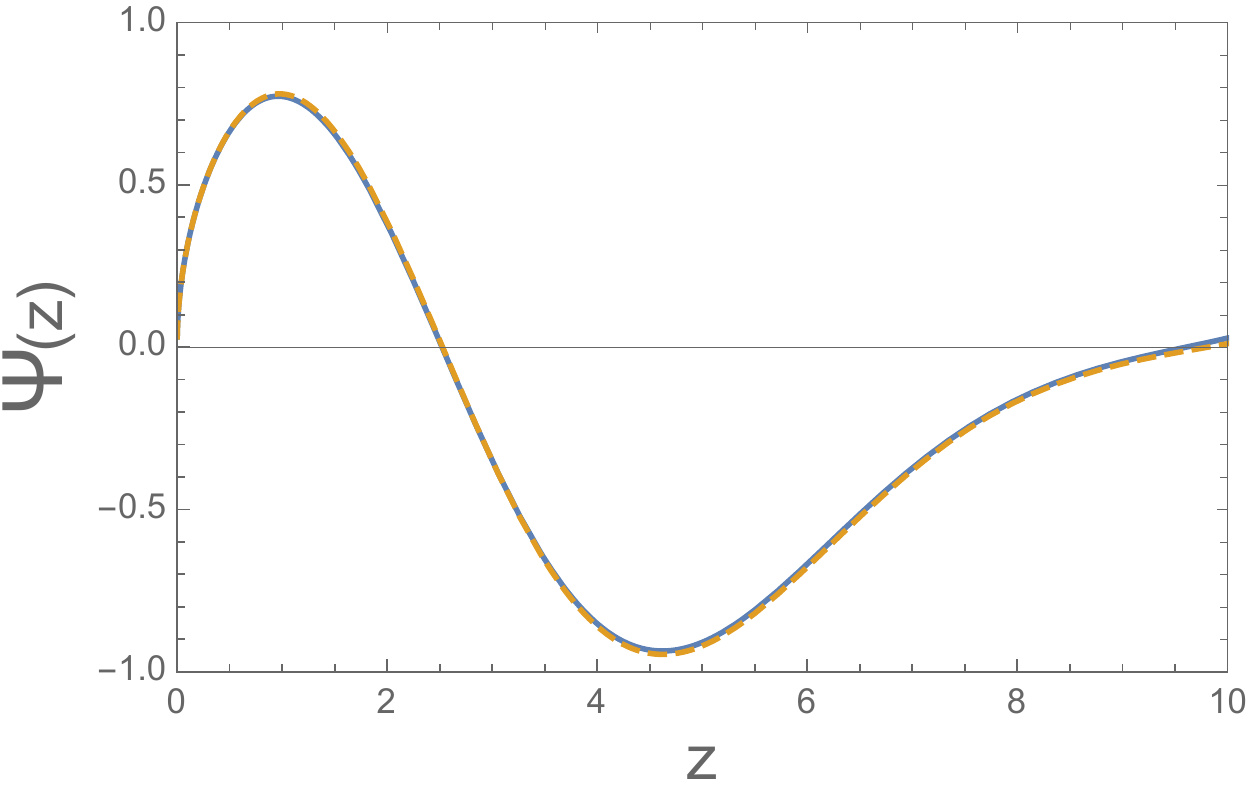}
\caption{The wave functions of the $\eta$ (left) and $\eta^{\prime}$ 
(right) with the new dilaton (solid), old dilaton (dashed). The 
new dilaton has been calculated with $\alpha=0.55$, 
$\delta=1.5235 $,$\gamma =1.0$ and $k =0.37/\sqrt{\alpha}$ GeV. The masses of the etas for 
the two dilaton calculations are identical $m_\eta =512$ MeV and $m_{\eta^{\prime}}=943$ 
MeV.}
\label{etanew}
\end{figure*}

\begin{table}[htb]
\begin{tabular}{| c | c | c | c | c | c | c|}
\hline
& $\pi^0$ & & $\pi(1300)$ & & & $\pi(1800)$    \\ \hline
PDG & $ 134.9768\pm 0.0005$ &  & $1300\pm100$ &  & & $1819\pm 10$  \\ \hline
This work &   $ 135  $    &$943 \pm 111$&       $ 1231  \pm 133$& $  1463 \pm151$ & $1663 \pm168$ & $1842\pm183$\\ \hline
\end{tabular}
\caption{ We show the experimental result for the $\pi$ masses given by the PDG particle listings~\cite{Tanabashi:2018oca,Zyla:2020zbs} together with the results of our calculation. We have used the $\pi^0$ mass to fix the value of $\delta=1.5235$. Our errors are again associated with the error in $\alpha=0.55 \pm 0.04$. The masses are as always in MeV.}
\label{piontab}
\end{table}

\begin{figure*}[htb]
\centering
\includegraphics[scale=0.65]{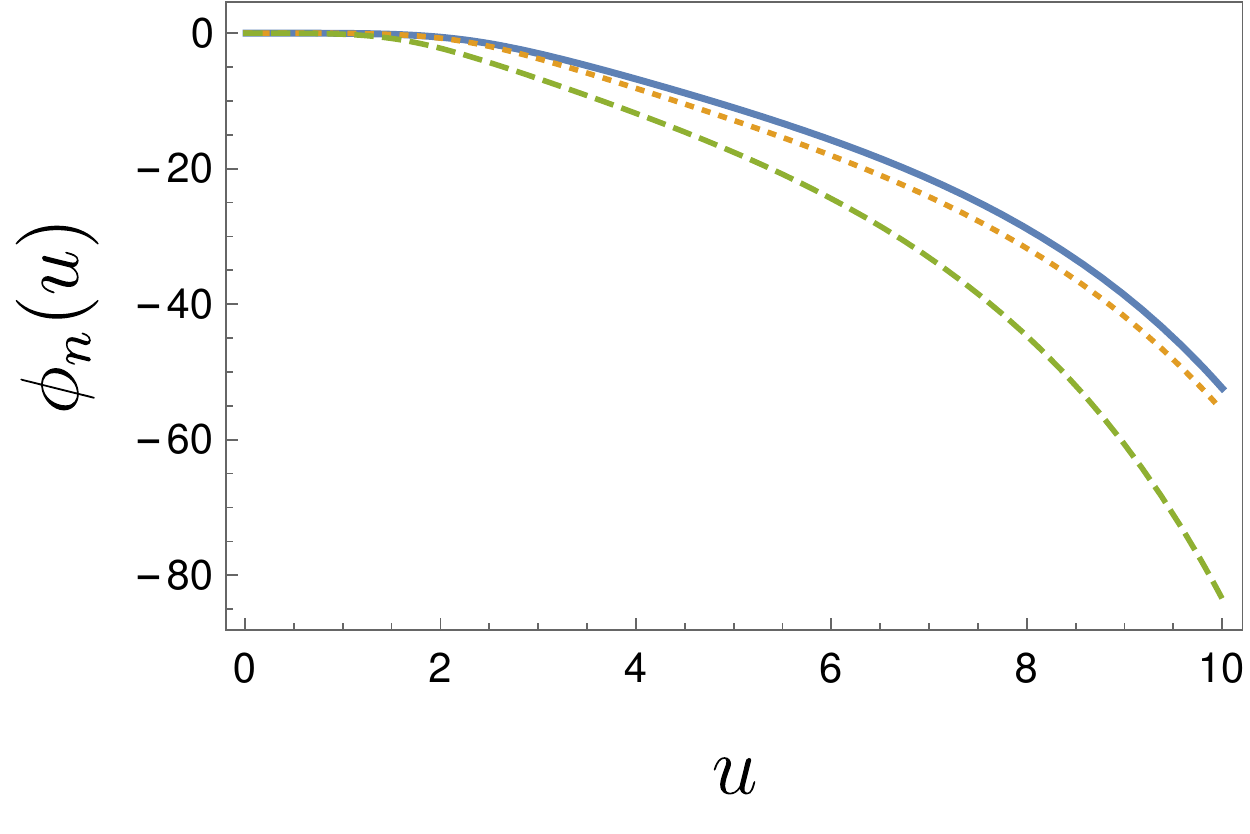}
\caption{The dilaton addition $\phi_n$ for the pseudo-scalar systems, 
see Appendix \ref{appA}. Full fine: the calculation for the pion with  the 
initial dilaton  $\phi_0$ of Eq. (\ref{49}). Dotted line: results obtained for 
the $\eta$ meson with the initial dilaton $\phi_0=\beta_s k^2 z^2$. Dashed 
line: 
the same for the scalar meson. Here $u=\alpha k^2 z^2$.}
\label{piondil}
\end{figure*}

\section{Conclusions}
\label{conc}

In the present investigation, a phenomenological analysis of the glueball and 
meson 
spectra within the GSW model is provided. This approach
 is based on the assumption that the lowest scalar glueball is associated to a 
graviton propagating in a deformed $AdS_5$ space. We saw in the past 
that the metric is fundamental in providing a good representation of the 
experimental data with only one parameter, and in so doing we determined the 
energy scale $k$ of the model. No approximation to the metric leads to a 
reasonable result, the graviton requires of the full power of the metric to 
produce the adequate experimental slope  and, in turn, to describe the correspondence to the confinement mechanism. The 
next step was to fit the scalar mesons within the same model. In this case the 
$AdS_5$ mass is negative and therefore the corresponding mode potential does not 
bind. In order to make the potential confining we had to truncate the 
exponential at third term. After doing so we obtain an excellent fit to 
the { light meson}
spectrum with only one additional parameter associated with the strength of the 
metric $\alpha$. With these two parameters fixed we have proceeded to describe 
the whole glueball and {light} meson spectrum. 
In all fits of glueballs, use  has been made of the full metric since the corresponding  
$AdS_5$ masses are positive. On the other hand, for all mesons, except for the 
$\rho$, whose $AdS_5$ mass is zero, we had to truncate the 
exponential metric at the third term to get a binding potential.   
With this procedure we have reproduced quite well the mass  spectra of
 the $\rho$, the $a_1$, the $\eta$ and the pion. 
While for the ground state pion a modification of the dilaton profile function
 is required to 
implement the chiral symmetry breaking, for all the other  hadrons, the 
masses have been calculated without any fit of the model parameters, which are 
only two, $k$ and $\alpha$, which were fixed by the scalar glueball and light 
scalar meson spectra. Such feature underlines the predictive power of the proposed model in 
describing the hadron masses. 
{We recall that the model is also able to fit well the  heavy scalar meson spectra.
We can conclude after this phenomenological analysis that the GSW
model provides a good description of the spectra of the axial and vector mesons, high spin glueballs and 
pseudo-scalar mesons and even heavy mesons with very few parameters.  Moreover, the model also  predicts the existence 
of further states not yet observed probably due  to the present experimental accuracy. }

{The success of this phenomenological meson potential has led us to
investigate how  the exponential metric is related to it. We have proven that a 
modification in the dilaton field is able to generate the phenomenological 
potential from the full metric. The proof is based on the construction of a 
differential equation for the dilaton field which relates the initial full 
potential with the phenomenological potential. We have shown that in our case, 
for all the mesons studied, the differential equation is solvable and moreover 
the new dilaton introduces no new parameters since it is defined exclusively by 
the metric parameters and the corresponding AdS mass. This new dilaton 
represents additional QCD interactions modifying, in the case of the mesons, 
the 
confining mechanism of glueballs.}

{We have compared the graviton solution for the glueballs, which described 
the scalar and tensor glueball spectrum with the $J=0$ and $J=2$  glueball field 
solutions.  We have seen that the degeneracy between scalars and glueballs of 
the former is instrumental in describing the spectra with only one energy scale. 
The field solutions require different energy scales for the $J=0$ and $J=2$ 
solutions since the one that fits the scalars leads to extremely heavy tensors, 
and the one that fits the tensor to extremely light scalars.  The graviton 
seems to be a necessary ingredient of AdS/QCD and the implications of this fact on QCD have 
to be understood. For the higher $J$ glueballs the field approximation is adequate and  it has allowed us to calculate successfully  the Regge trajectories of the even and odd high spin glueballs.}
The scalar glueball and the pion escape this 
scheme. The former requires a graviton 
propagating in a deformed $AdS$ space 
and the latter a sophisticated dilaton. 
Clearly this might be associated in QCD with the fact that the ground state 
scalar glueball is associated  with the $\sigma $ particle in some schemes and 
the pion with the Goldstone boson of spontaneously broken chiral symmetry.

One should notice that the GSW model, like other phenomenological 
approaches based on the  
the AdS/CFT correspondence, is realized in the large $N$ approximation. 
Therefore,
one might expect higher order corrections to be required for precision 
calculations,  which are beyond the aim of the present investigation. 
Finally let 
us conclude by remarking the surprising capability of the model in reproducing 
basic features of many different hadronic systems without invoking  a large 
number of parameters and therefore unveiling a relevant predicting power 
that could be used in future analyses.

\section{Acknowledgements}

This
work was supported, in part by the STRONG-2020 project of the European Unions Horizon 2020 research and
innovation programme under grant agreement No 824093. This work was also supported in part by MICINN, AEI and UE FEDER under contract PID2019-105439-GB-C21.

\appendix
\section{The dilaton differential equation for the scalar and pseudo-scalar fields}
\label{appA}

In this appendix details on the differential equation (\ref{difeqs}) will be 
provided. In particular, since for the pion the initial dilaton ($\phi$) must 
be properly  chosen in order to introduce $\chi$sb into the model, here we 
provide the general differential equation that the dilaton  addition
($\phi_n$)  must satisfy  to generate a binding potential where 
the 
metric effects are encoded in the truncated expansion of $e^{\alpha k^2 z^2}$. 
Let us start again with the full general action for a scalar field:

\begin{align}
\bar S = \int d^5x ~\sqrt{-g} e^{- \phi(z) +\frac{3}{2} \alpha k^2z^2
 -\phi_n(z)  } \Big[  g^{MN} 
\partial_M S(x) \partial_N S(x)+ e^{\alpha k^2 z^2} M_5^2 R^2S^2(x) \Big]~,
\end{align}
where we recall that for $\phi(z)= k^2 z^2 \beta_s$ and $\beta_s = 1+ 
\frac{3}{2} \alpha$ we get the usual result Eq. (\ref{mod1}) and of Ref. 
\cite{Rinaldi:2020ssz}. From the Euler-Lagrange equation and by properly 
choosing a functional form the field $S(x)$, a Schr\"odinger equation can be 
obtained,

\begin{align}
 -\psi''(z)+ V_s(z) \psi(z) =M^2 \psi(z)~,
\end{align}
where the  potential  $V_s(z)=\tilde V_s(z)/(4 z^2)$ and,

\begin{align}
\label{pott}
 \tilde V_s(z) &= 4 e^{\alpha k^2 z^2} M_ 5^2+3 \Big[5+\alpha k^2 
z^2 \
(-4+3 \alpha k^2 z^2)\Big]+
\\
\nonumber
&+ z \Big \{z \phi'(z)^2+(6-6 \alpha k^2 z^2) \
\phi_n'(z)+z \phi_n'(z)^2+\phi'(z) \big[6-6 \alpha k^2 z^2+2 z \phi_n'(z) 
\big]-2 z \big[\phi''(z)+\phi_n''(z) \big] \Big\}~.
\end{align}
The equivalent quantity obtained for $\phi_n=0$ and $\phi = 
k^2 z^2 (1+ \frac{3}{2} \alpha)$ becomes ~\cite{Rinaldi:2020ssz},

\begin{align}
 \tilde V_s^o(z) =  15 + 4 M_5^2 R^2 e^{\alpha k^2 z^2}+ 8 k^2 z^2+ 4 k^4 z^4~,
\end{align}
Moreover, the binding potential, 
needed to reproduce the scalar and pseudo-scalar spectra previously discussed, 
must be obtained by setting $\phi_n=0$ and 
by expanding the exponential term $e^{\alpha k^2 z^2} $ up to the 
third term:

\begin{align}
\label{potta}
 \tilde V_s^a(z) &= 4(1 + \alpha k^2 z^2+ \frac{1}{2}\alpha^2 k^4 z^4) M_ 5^2+3 
\Big[5+\alpha k^2 
z^2 \
(-4+3 \alpha k^2 z^2)\Big]+
\\
\nonumber
&+ z \Big \{\phi'(z) \big[6-6 \alpha k^2 z^2+2 z  
\big]-2 z \phi''(z) \Big\}~.
\end{align}
 Therefore, the differential equation, that the correction dilaton $\phi_n$ 
must satisfy  to move from the general potential Eq. (\ref{pott}) to the 
expression Eq. (\ref{potta}) is,

\begin{align}
  \tilde V_s(z)- \tilde V_s^a(z)=4 M_5^2 R^2 \Big[ e^{\alpha k^2 z^2} - 1 - 
\alpha k^2 z^2 - \frac{1}{2}\alpha^2 k^4 z^4\Big]+z \Big[\phi_n'(z) \big(6-6 
\alpha k^2 z^2+2 z \phi'(z)+z  \phi_n'(z) \Big)-2 z \phi_n''(z) \Big]=0~.
\end{align}
The above equation can be simplified to:

\begin{align}
 - \frac{\phi''_n(z)}{2}+ 
\frac{\phi'_n(z)^2}{4}+\phi'_n(z) \left( \frac{3}{2z}-\frac{3}{2}\alpha k^2 
z+\frac{1}{2}\phi'_n(z) \right)+\frac{M_5^2 R^2}{z^2} \left(e^{\alpha k^2 
z^2}-1-\alpha k^2 z^2-\frac{1}{2}\alpha^2 k^4 z^4  \right)=0~.
\end{align}

  As one can see this equation directly depend on the old initial dilaton 
$\phi$, therefore such a procedure can be applied in the scalar and 
pseudo-scalar ($\eta$ and $\pi$) mesons. In the case of scalar meson and $\eta$, 
$\phi(z) = k^2 z^2(1+ \frac{3}{2} \alpha)$ one gets:

\begin{align}
\label{dis}
 - \frac{\phi''_n(z)}{2}+ 
\frac{\phi'_n(z)^2}{4}+\phi'_n(z) \left( \frac{3}{2z}+ k^2 
z \right)+\frac{M_5^2 R^2}{z^2} \left(e^{\alpha k^2 
z^2}-1-\alpha k^2 z^2-\frac{1}{2}\alpha^2 k^4 z^4  \right)=0~.
\end{align}

The numerical solution has been used to fit $\phi_n$ as a polynomial function 
of $u=\alpha k^2 z^2$:

\begin{align}
 \phi_n(u) \sim 0.507286 - 0.035493 u^{1.5}-0.800325 
u^2+0.0052429 u^4-0.0000556475 u^6~.
\end{align}

\section{The dilaton differential equation for the vector field}
\label{appB}
The same procedure can be extended to vector fields. In this case 
dilatons describing chiral symmetry breaking are not considered in the analysis. We show 
the differential equation for $\phi_n$ given $\phi_0= \beta_\rho k^2 z^2$ where 
$\beta_\rho =1+ \frac{1}{2}\alpha $. 

In this case,

\begin{align}
 \bar S_V = -\frac{1}{2} \int d^5x \sqrt{-g} e^{-k^2z^2-\phi_n} 
\left[\frac{1}{2} g^{MP} g^{QN} F_{MN}F^{PQ}+ M_5^2 R^2g^{PM} A_P A_M e^{\alpha 
k^2 z^2} \right]~.
\end{align}
From the EoM one can derive the potential:

\begin{align}
 V_v(z)&= \left[\frac{B'(z)^2}{4}-\frac{B''(z)}{2}+\frac{M_5^2 R^2}{z^2} 
e^{\alpha k^2 z^2}  \right]
\\
\nonumber
&=-\frac{\phi_n''(z)}{2}+ \frac{\phi_n'(z)^2}{4}+\phi_n'(z) 
\left(\frac{1}{2z}-k^2 z \right)+ k^4 z^2+ \frac{3}{4 z^2} +\frac{M_5^2 
R^2}{z^2} 
e^{\alpha k^2 z^2}~,
\end{align}
where here $B(z)=\log(z)+\phi_n(z)-k^2 z^2$. Also in this case, the 
approximated potential $V_v^a$ is obtained for $\phi_n=0$ and by expanding 
$e^{\alpha k^2 z^2}$ up to the third term. The 
phenomenological potential is,

\begin{align}
 V_v^a(z)= k^4 z^2+\frac{3}{4z^2}+ \frac{M_5^2 R^2}{z^2}(1 +\alpha k^2 z^2 + 
\frac{1}{2} \alpha^2 k^4 z^4)~,
\end{align}
 therefore the differential equation reads,

\begin{align}
\label{div}
  V_v(z)- V_v^a(z)= -\frac{\phi_n''(z)}{2}+ \frac{\phi_n'(z)^2}{4}+\phi_n'(z) 
\left(\frac{1}{2z}-k^2 z \right) +\frac{M_5^2 R^2}{z^2} \left(
e^{\alpha k^2 z^2}-1-\alpha k^2 z^2 - \frac{1}{2}\alpha^2 k^4 z^4 \right)=0.
\end{align}

\section{The general dilaton differential equation for $\phi(z)=\beta k^2 z^2$}
\label{appC}
Due to the similarities between Eqs. (\ref{dis}, \ref{div}), we show here that 
the although the dilaton expression would explicitly depend on the kind of 
meson (scalar or vector), no free parameter are involved and the 
differential equation can be written through a general expression:

\begin{align}
 -\frac{\phi_{n,I}^{''}(z)}{2}+ \frac{\phi_{n,I}'(z)^2}{4}+\phi_{n,I}'(z) 
\left(\frac{A_I}{2z}+ B_Ik^2 z \right) +\frac{M_5^2 R^2}{z^2} \left(
e^{\alpha k^2 z^2}-1-\alpha k^2 z^2 - \frac{1}{2}\alpha^2 k^4 z^4 \right)=0~,
\end{align}
where for the scalar we have $I=s,~ A_s=3,~ B_s=1$ and for the vector $I=v,~ 
A_v=1,~ B_v=-1$.

\newpage

\bibliographystyle{apsrev4-1}
%\bibliography{/home/matteo/Scrivania/Lavoro/latex/REFERENZE/bib_glueball3}
\bibliography{GSWPRD20212.bib}

%\begin{thebibliography}{4}

%\end{thebibliography}

\appendix*

\end{document}